\documentclass[12pt]{article} 
\usepackage{epsfig}
\usepackage{a4}
\usepackage{latexsym}
\usepackage{cite}

\usepackage{pslatex}
\usepackage[latin1]{inputenc}
\usepackage[T1]{fontenc}

\newcommand{\lsim}{\raisebox{-0.07cm}{$\:\:\stackrel{<}{{\scriptstyle
 \sim}}\:\: $} }

\newcommand{\hspn}{{\hspace{-5mm}}}
\newcommand{\hspp}{{\hspace{3mm}}}
\newcommand{\beq}{\begin{equation}}
\newcommand{\eeq}{\end{equation}}
\newcommand{\bea}{\begin{eqnarray}}
\newcommand{\eea}{\end{eqnarray}}
\newcommand{\nn}{\nonumber}
\newcommand{\MSb}{$\overline{\mbox{MS}}$}
\newcommand{\as}{\alpha_{\rm s}}
\newcommand{\ar}{a_{\rm s}}
\newcommand{\ra}{\rightarrow}
\newcommand{\ep}{\epsilon}
\newcommand{\eps}{\epsilon^{\:\!2}}
\newcommand{\ept}{\epsilon^{\:\!3}}
\newcommand{\epf}{\epsilon^{\:\!4}}
\newcommand{\Nbar}{\bar{N}}

\def\fr#1#2{{\textstyle{{#1}\over{#2}}}}
\def\FR#1#2{{\displaystyle{{#1}\over{#2}}}}

\begin{document}

\setlength{\parskip}{0.2cm}
\setlength{\baselineskip}{0.53cm}

\def\Qs{{Q^{\, 2}}}
\def\ca{{C_A}}
\def\cas{{C^{\: 2}_A}}
\def\cath{{C^{\: 3}_A}}
\def\cafo{{C^{\: 4}_A}}
\def\cf{{C_F}}
\def\cfs{{C^{\: 2}_F}}
\def\cfth{{C^{\: 3}_F}}
\def\nf{{n^{}_{\! f}}}
\def\nfs{{n^{\,2}_{\! f}}}
\def\nft{{n^{\,3}_{\! f}}}

\def\z#1{{\zeta_{#1}}}

\begin{titlepage}

\noindent
LTH 920 \hspace*{\fill} August 2011\\
LPN11-44 \\
\vspace{2.5cm}
\begin{center}
\Large
{\bf Resummation of small-x double logarithms in QCD$\,$:\\[1mm]
Semi-inclusive electron-positron annihilation}\\
\vspace{2cm}
\Large
A. Vogt\\
\vspace{1cm}
\normalsize
{\it Department of Mathematical Sciences, University of Liverpool \\
\vspace{1mm}
Liverpool L69 3BX, United Kingdom}\\[2.5cm]
\vfill
\large
{\bf Abstract}
\vspace{-0.2cm}
\end{center}
We have derived the coefficients of the highest three $1\!/x\,$-enhanced 
small-$x$ logarithms of all timelike splitting functions and the coefficient 
functions for the transverse fragmentation function in one-particle inclusive 
$e^{\:\!+}e^{\:\!-}$ annihilation at (in principle) all orders in massless 
perturbative QCD. \linebreak
For the longitudinal fragmentation function we present the respective two 
highest contributions.
These results have been obtained from KLN-related decompositions of the
unfactorized fragmentation functions in dimensional regularization and their 
structure imposed by the mass-factorization theorem.
The resummation is found to completely remove the huge small-$x$ spikes present
in the fixed-order results, allowing for stable results down to very small 
values of the momentum fraction and scaling variable~$x$.  
Our calculations can be extended to (at least) the corresponding
$\as^{\,n} \ln^{\,2n -\ell\!} x$ contributions to the above quantities and their
counterparts in deep-inelastic scattering.
   
\vspace{0.5cm}
\end{titlepage}
%

\section{Introduction}
One-hadron inclusive electron-positron annihilation, $\,e^{\:\!+}e^{\:\!-}\ra\: 
\gamma\,,\,Z \,\ra\, h + X$ where $h$ denotes the observed hadron (or a sum 
over all charged hadron species) and $X$ any inclusive hadronic final state, 
is an important benchmark process in perturbative QCD which has been measured 
accurately over a wide range of centre-of-mass (CM) energies $\!\sqrt{s}$
\cite{PDG10}.
The results provide crucial inputs for fit determinations of the fragmentation
distributions (or parton fragmentation functions) $\,D_{\! p}^{\,h}(x,\Qs)$,
see Refs.~\cite{FrgFits1,FrgFits2,FrgFits3},
where $x$ represents the fraction of the momentum of the final-state parton~$p$
transferred to the outgoing hadron $h$ and $\Qs$ is a hard scale, for instance
the squared four-momentum $q$ of the timelike virtual photon or $Z$-boson in 
the above semi-inclusive annihilation (SIA) process, $\Qs = q^{\:\!2} = s$.  
SIA data have also provided constraints on the strong coupling constant $\as$ 
\cite{FrgAs}.

The theoretical description of semi-inclusive $\,e^{\:\!+}e^{\:\!-}$ 
annihilation is analogous to that of electron-hadron deep-inelastic scattering 
(DIS), $\,ep \,\ra\, e + X$, via the exchange of a (spacelike) virtual photon 
or $Z$-boson. 
The SIA differential cross section can be written in terms of transverse ($T$),
longitudinal ($L$) and asymmetric ($A$) fragmentation functions 
(timelike structure functions) \cite{NasWeb94},
\beq
\label{FrgFct}
  {1 \over \sigma_0^{}}\, {d^{\:\!2} \sigma \over dx\: d\!\cos \theta}
  \;\; = \;\; 
               \FR38\, (1 + \cos^2 \theta) \; F_T^{\:\! h}(x,\Qs)
         \:+\: \FR34\: \sin^2 \theta \; F_L^{\:\! h}(x,\Qs)
         \:+\: \FR34\: \cos \theta \; F_A^{\:\! h}(x,\Qs) \:\: .
\eeq
Here $x = 2E_h/\sqrt{s}\le 1$ and $\theta$ are the scaled energy of the hadron 
$h$ and its angle relative to the electron beam, respectively, in the CM frame;
and for photon exchange $\,\sigma_0^{} = n_c\: 4\pi\,\alpha^{2\!}/3s$ is the 
total cross section for Bhabha scattering times the number of colours $n_{c\,}$.
Disregarding corrections suppressed by inverse powers of $Q$, the fragmentation
functions are related to the fragmentation distributions by
 
\vspace*{-7mm}
\beq
\label{FaCD} \quad
  F_a^{\:\! h}(x,\Qs) \;\; = \; \sum_{p \,=\, \rm q,\,\bar{q},\,g} \;
  \int_x^1 {dz \over z} \; c^{}_{a,p} \!\left( z,\as (\Qs) \right)
  \:  D_{\! p}^{\,h} \Big( {x \over z},\,\Qs \Big)
  \:\: .
\eeq
The coefficient functions $c^{}_{a,p}$ in Eq.~(\ref{FaCD}) are known to order 
$\as^{\,2}$ \cite{NLOcoeff1,NLOcoeff2,RvN2loop,MM06}, see also 
Ref.~\cite{MMV06}, i.e., to the next-to-next-to-leading order (NNLO) for $F_T$ 
and $F_A$ and to the next-to-leading order (NLO) for $F_L$ which vanishes for 
$\as=0$. 
Here and throughout this article we identify, without loss of information,
the \MSb\ renormalization and mass factorization scales with the physical hard 
scale~$\Qs$.

The scale dependence of the (process-independent) final-state fragmentation 
distributions is analogous to that of the initial-state parton distributions 
and given~by
\beq
\label{Devol}
  {d \over d \ln \Qs} \; D_{i}^{\,h}(x,\Qs) \;\; = \;
  \sum_{j \,=\, \rm q,\,\bar{q},\,g} \;
  \int_x^1 {dz \over z} \; P^{\,T}_{\! ji} \left( z,\as (\Qs) \right)
  \:  D_{\! j}^{\,h} \Big( {x \over z},\, \Qs \Big) \:\: .
\eeq
The (timelike) splitting functions $\,P^{\,T}_{ji}\,$ can be expanded in 
powers of $\ar \equiv \as(\Qs)/ (4\pi)$,
\beq
\label{PTexp}
  P^{\,T}_{\! ji} \left( x,\as (\Qs) \right) \:\: = \:\:
  \ar \, P_{\! ji}^{\,(0)\,T}(x) \: + \: \ar^{\:\!2} \, P_{\! ji}^{\,(1)\,T}(x)
  \: +\: \ar^{\:\!3} P_{\! ji}^{\,(2)\,T}(x) \: +\: \ldots \;\; .
\eeq
The leading-order (LO) and NLO contributions $P^{(0)\,T}$ and $P^{(1)\,T}$
to Eq.~(\ref{PTexp}) have been known for a long time 
\cite{PT1loop,CFP80,KKST80,FKL81,MOKK01}. 
A direct calculation of the NNLO corrections $P^{(2)\,T}$ has not been 
performed so far. However, an indirect determination \cite{MMV06,MV2}, using 
non-trivial relations to the spacelike DIS case 
\cite{MVV34} and the supersymmetric limit 
\cite{CFP80,FKL81,AF81,SV96,BRvN00,DMS05,BK06} has been completed recently 
\cite{AMV1} up to a minor caveat, which is not relevant in the present context,
concerning the quark-gluon splitting. 

\pagebreak

Eq.~(\ref{PTexp}) and the corresponding fixed-order approximations to the 
coefficient functions (see below) are adequate except for $1\!-\!x \ll 1$ and 
$x \ll 1$, where higher-order corrections generally include double logarithms 
which can spoil the perturbative expansions. Here we focus on the small-$x$
case, where the leading contributions to the N$^n$LO splitting functions are
of the form
\beq
\label{PgjLL}
  P_{ji}^{\,(n)\,T\!}(x) \;\;=\;\; 
  \delta_{jg}\, a_{i}^{(n)}\: \FR{1}{x}\, \ln^{\,2n\!} x \;+\; \ldots 
  \quad , \quad 
  a_{\rm q}^{\,(n)} \;=\; \FR{\cf}{\ca}\: a_{\rm g}^{\,(n)} 
\eeq
where $\delta_{\,ij}$ is the Kronecker symbol, and $C_A$ and $C_F$ are the
standard SU(N) colour factors, with $C_A = n_c = 3$ and $C_F = 4/3$ in QCD. The 
coefficients $a_{i}^{\,(n)}$ and the corresponding subleading contributions lead
to corrections which are numerically far larger than the corresponding single-%
logarithmic enhancement of the analogous spacelike N$^n$LO splitting functions 
governing the DIS case \mbox{\cite{BFKL1,BFKL2,BFKL3,CH94,NL-BFKL}}; for $n=2$ 
see Figs.~1 of Refs.~\cite{MV2} and \cite{AMV1}. On the other hand, the 
all-order Mellin-space summation of the leading-logarithmic (LL) contributions 
(\ref{PgjLL}) leads to \cite{LLxto0}
\beq
\label{PLLsum}
  {C_A \over C_F}\: P_{\rm gq}^{\,T}(N,\as) \;=\;
  P_{\rm gg}^{\,T}(N,\as) \;=\;
  \FR{1}{4}\: (N-1) \left\{ \! 
    \left( 1 + \FR{32\:\! \ca \ar}{(N-1)^2}  \right)^{1/2}\! - 1
  \right\} \;\; + \;\; \mbox{ NLL terms} 
\eeq
which can be expanded to all orders in $x$-space via the standard Mellin 
transform
\beq
\label{Mlog}
  {\rm M} \left[ \,\FR{1}{x}\: \ln^{\,k} x \right](N) \;\: \equiv \;\:
  \int_0^1 \! dx\: x^{\,N-1} \: \FR{1}{x}\: \ln^{\,k\!} x \;\: = \;\:
  \frac { (-1)^{k\,} k! }{(N-1)^{k+1} } \:\;\; .
\eeq
Eq.~(\ref{PLLsum}) corresponds to a small and oscillating function in $x$-space,
suggesting that the small-$x$ enhancement of $P_{\!\rm gi}^{(1)T}(x,\as)$ and 
$P_{\!\rm gi}^{(2)T}(x,\as)$ -- which is negative in the former and positive in 
the latter case, see below -- is unphysical and can be removed by extending  
Eq.~(\ref{PLLsum}) to the next-to-leading logarithmic (NLL) and 
next-to-next-to-leading logarithmic (NNLL) small-$x$ accuracy. 
Even the former extension
has not been performed in the \MSb\ scheme so far, as the results of 
Ref.~\cite{MuellerNL} are not given in this scheme (and consequently do not 
agree with the NNLO next-to-leading logarithms of Refs.~\cite{MV2,AMV1}). 
For a detailed discussion see Ref.~\cite{ABKK11} where also the LL result for 
the \MSb\ transverse coefficient function $c_{T,\,\rm g}^{}$ corresponding to 
Eq.~(\ref{PLLsum}) has been derived.

In this article we employ constraints provided by the structure of the 
unfactorized fragmentation functions in dimensional regularization 
\cite{DimReg} and the all-order mass-factorization formula to derive
the coefficients of the respective highest three 
non-vanishing logarithms for all four timelike splitting functions 
$P_{\! ji}^{\,T}(x,\as)$, $i,\,j = {\rm q,\,g}$, as well as the corresponding 
coefficients for both coefficient functions for $F_T$, to all (in practice 
sixteen) orders in $\as$. The derivation of the second$/$third logarithms
is made possible by the NLO$/$NNLO fixed-order results; consequently only the
highest two logarithms can be resummed for the longitudinal fragmentation
function $F_L$. 

The remainder of this article is organized as follows: In Section 2 we describe
the theoretical framework used to perform the resummation and comment on the
calculations which were carried out using the latest version of {\sc Form} and
{\sc TForm} \cite{Form3,TForm}. 
The resummed splitting functions are written down and discussed in Section~3.
The corresponding results for the transverse and longitudinal coefficient
functions are presented in Sections 4 and 5, respectively. Our findings are
summarized in Section 6, which also provides a brief outlook to future 
applications and extensions.

\setcounter{equation}{0}
\section{Method and calculation}
The main quantities in our resummation are the unfactorized flavour-singlet 
partonic fragmentation functions in $D = 4 - 2\ep$ dimensions
\beq
\label{TaCZ}
  \widehat{F}_{a, k}(N,\ar,\ep) \;\;=\;\; 
  \widetilde{C}_{a, i}(N,\ar,\ep) \; Z_{\,ik}^{\,T}(N,\ar,\ep) 
\eeq
where the summation over $i = \rm q,g$ and the \MSb\ removal of $(4\pi)^\ep$
and $\gamma_e$ factors \cite{BBDM78} are understood.
The $D$-dimensional coefficient functions $\widetilde{C}_{a, i}$ include all 
non-negative powers of $\ep$ in Eq.~(\ref{TaCZ}),
\beq
\label{Ctilde}
 \widetilde{C}_{a, i}(N,\ar,\ep) \;=\;
 \delta_{\,a\:\!T}\, \delta_{\,i\:\!\rm q} \:+\:
 \delta_{\,a\:\!\phi}\, \delta_{\,i\:\!\rm g}
 \:+\: \sum_{\ell=1}^{\infty}\, \ar^{\,\ell}\, \sum_{k=0}^{\infty}\: \ep^{\,k}
 c_{a, i}^{\,(\ell,k)}(N) \; .
\eeq
Besides the fragmentation functions $F_T$ and $F_L$ of Eq.~(\ref{FrgFct}) 
-- $F_A$ is a non-singlet quantity without $1/x$ terms -- we consider SIA with
an intermediate scalar $\phi$ coupling directly only to gluons via an
additional term $\phi\,G^{\,\mu\nu\!}G_{\mu\nu}$ in the Lagrangian, where 
$G^{\,\mu\nu}$ represents the gluon field strength tensor.
Such an interaction, suggested as a QCD trick in Ref.~\cite{FP82}, does 
occur in the Standard Model for the Higgs boson in the limit of a very heavy 
top quark \cite{HGGeff}. The NLO and NNLO quark and gluon coefficient functions
for the resulting fragmentation function $F_\phi$ have been presented in 
Ref.~\cite{AMV1}.

The final-state transition functions $Z_{\,ik}^{\,T}$ collecting all negative 
powers of $\ep$ are related to the matrix of the splitting functions in 
Eq.~(\ref{PTexp}) by
\beq
\label{PofZ}
  P^{\:T} \;\:=\;\: \frac{d\:\! Z^{\,T} }{d\ln \Qs }\,
                    \left( Z^{\,T} \right)^{-1}
          \;\:=\;\: \beta_D (\ar) \: \frac{d\:\! Z^{\,T}}{d\ar}\,
                    \left( Z^{\,T} \right)^{-1} \
\eeq
 
\vspace*{-5mm}
\noindent
with
\beq
\label{PTmat}
  - \,\gamma  \;\: \equiv \;\:
  P^{\:T} \;=\;\: \left( \begin{array}{cc}
         P_{\rm qq}^{\,T} & P_{\rm gq}^{\,T} \\[2mm]
         P_{\rm qg}^{\,T} & P_{\rm gg}^{\,T}
   \end{array} \right)
  \quad \mbox{ and } \quad
  \beta_D (\ar) \;\:=\;
   -\,\ep\, \ar - \beta_0 \,\ar^2 - \beta_1 \,\ar^3 - \: \ldots \;\; .
\eeq
As we shall see from the next equation, only the leading coefficient of the 
four-dimensional beta function of QCD, $\beta_0 = 11/3\;\ca - 2/3\;\nf\,$ 
\cite{beta0} 
where $\nf$ stands for the number of effectively massless quark flavours, 
enters the resummation of the highest three small-$x$ logarithms.

Eq.~(\ref{PofZ}) can be solved for $Z$ order by order in $\as$. Suppressing
all functional dependences, as already done for most quantities in the previous
two equations, the first four orders read
\bea
\label{ZofP}
  Z^{\,T}\!\! &\!=\!& 1 \:+\: \ar \:{1 \over \ep}\: \gamma_0^{}
  \:+\: \ar^{\,2} \bigg\{ \, {1 \over 2\:\!\eps}\,
       \left( \gamma_0^{} - \beta_0 \right) \gamma_0^{}
       \:+\: {1 \over 2\:\!\ep}\: \gamma_1^{} \bigg\}
  \nn \\[0.5mm] & & \mbox{\hspn}
  \:+\: \ar^{\,3} \bigg\{ \, {1 \over 6\:\!\ept}
       \left( \gamma_0^{} - \beta_0 \right)
       \left( \gamma_0^{} - 2\:\!\beta_0 \right) \gamma_0^{}
       \:+\: {1 \over 6\:\!\eps}
       \Big[ \left( \gamma_0^{} - 2\:\!\beta_0 \right)
       \gamma_1^{} + \left( \gamma_1^{} - \beta_1 \right)
       2\:\!\gamma_0^{} \Big] + {1 \over 3\:\!\ep}\: \gamma_2^{}
       \bigg\}
  \nn \\[0.5mm] & & \mbox{\hspn} 
  \:+\: \ar^{\,4} \bigg\{ \, {1 \over 24\:\!\epf}
       \left( \gamma_0^{} - \beta_0 \right)
       \left( \gamma_0^{} - 2\:\!\beta_0 \right) 
       \left( \gamma_0^{} - 3\:\!\beta_0 \right) \gamma_0^{} 
  \nn \\[0.5mm] & & \mbox{\hspp} 
       \:+\: {1 \over 24\:\!\ept} \Big[ 
         \left( \gamma_0^{} - 2\:\!\beta_0 \right) 
         \left( \gamma_0^{} - 3\:\!\beta_0 \right) \gamma_1^{}
       + \left( \gamma_0^{} - 3\:\!\beta_0 \right)
         \left( \gamma_1^{} - \beta_1 \right) 2 \gamma_0^{}
       + \left( \gamma_1^{} - 2\:\!\beta_1 \right)
         \left( \gamma_0^{} - \beta_0 \right) 3 \gamma_0^{}
       \Big]
  \nn \\[0.5mm] & & \mbox{\hspp}
       \:+\: {1 \over 24\:\!\eps} \Big[
         \left( \gamma_0^{} - 3\:\!\beta_0 \right) 2 \gamma_2^{}
       + \left( \gamma_1^{} - 2\:\!\beta_1 \right) 3 \gamma_1^{}
       + \left( \gamma_2^{} - \beta_2 \right) 6 \gamma_0^{}
       \Big]
       \:+\: {1 \over 4\:\!\ep}\: \gamma_3^{}
       \bigg\}
  \:\;+\;\; \ldots \;\; .
\eea
The corresponding higher-order contributions have been generated in {\sc Form}
to order $\as^{\,16}$.
It is clear from these results that the N$^n$LO corrections, i.e., the 
splitting functions up to $\gamma_{\:\!n}^{} \equiv \gamma^{\,(n)}$ defined 
analogous to Eq.~(\ref{PTexp}) together with $\beta_0, \:\ldots,\: \beta_n$, 
determine the highest $\,n\!+\!1\,$ powers of $1/\ep$ in Eq.~(\ref{ZofP}) at 
all orders in $\as$. 
Keeping in mind $\gamma_{\:\!n}^{} \propto 1/(N-1)^{2n+1}$, one notices that
$\beta_0$ and $\beta_0^{\,2}$ enter at NLL and NNLL small-$x$ accuracy only.
Moreover $\beta_1$ is suppressed by three powers in $1/(N-1)$ relative to
$\gamma_1^{}$; hence this coefficient contributes only at the level of the 
fourth logarithms, i.e., beyond our present accuracy.

The same considerations apply to the unfactorized structure functions 
(\ref{TaCZ}), which at N$^n$LO require the coefficients $c_{a, i}^{\,(\ell,k)}$
with $\,\ell+k \,\leq\, n\,$ for $F_T$ and $F_\phi$, and $\,\ell+k \,\leq\, 
n+1\,$ for $F_L$ in Eq.~(\ref{Ctilde}). 
For the convenience of the reader we collect the coefficient function
results which form the input of the small-$x$ resummation discussed below.
The expansions about $\Nbar \equiv N-1 = 0$ for $F_T$ read
\bea
  c_{T, \rm q}^{\,(1,0)} &\!=\!& \cf + {\cal O}(\Nbar) \;\; , \quad
  c_{T, \rm q}^{\,(1,1)} \:\:=\:\: {\cal O}(\Nbar^{\,0}) \;\; ,
\nn \\[1mm]
  c_{T, \rm q}^{\,(2,0)} &\!\cong\!& 
          \fr{64}{3}\: \cf\nf\, \Nbar^{\,-3}
    \,+\, \fr{16}{3}\: \cf\nf\, \Nbar^{\,-2} 
    \,-\, \fr{80}{27}\: \cf\nf\, \Nbar^{\,-1}
\label{cTq12}
\eea
and
\bea
  c_{T, \rm g}^{\,(1,0)} &\!\cong\!& 
      -\, 8\,\cf\, \Nbar^{\,-2}
    \,-\, 4\,\cf\, \Nbar^{\,-1}
    \,+\, ( \fr{27}{2} - 4\,\z2 )\,\cf\, \Nbar^{\,0}
    \; ,
\nn \\[1mm]
  c_{T, \rm g}^{\,(1,1)} &\!\cong\!&
      -\, 16\,\cf\, \Nbar^{\,-3}
    \,-\, 8\,\cf\, \Nbar^{\,-2}
    \,-\, ( 12 - 6\,\z2 )\,\cf\, \Nbar^{\,-1}
    \; ,
\nn \\[1mm]
  c_{T, \rm g}^{\,(2,0)} &\!\cong\!&
          160\,\cf\ca\, \Nbar^{\,-4}
    \,-\, \fr{232}{3}\: \cf\ca\, \Nbar^{\,-3}
    \,-\, \left[ \left( \fr{248}{3} + 16\,\z2 \right)\, \cf\ca 
          + 8\,\cfs \,\right] \, \Nbar^{\,-2}
    \; . \quad
\label{cTg12}
\eea
The corresponding results for $F_\phi$ are given by
\bea
  c_{\phi, \rm q}^{\,(1,0)} &\!=\!& -\,\fr{14}{3}\:\nf 
    \,+\, {\cal O}(\Nbar) \;\; , \quad
  c_{\phi, \rm q}^{\,(1,1)} \:\:=\:\: {\cal O}(\Nbar^{\,0}) \;\; ,
\nn \\[1mm]
  c_{\phi, \rm q}^{\,(2,0)} &\!\cong\!&
          \fr{64}{3}\: \ca\nf\, \Nbar^{\,-3}
    \,-\, \fr{16}{3}\: \ca\nf\, \Nbar^{\,-2}
    \,-\, \fr{296}{27}\: \ca\nf\, \Nbar^{\,-1}
\label{cPq12}
\eea
and
\bea
  c_{\phi, \rm g}^{\,(1,0)} &\!\cong\!&
      -\, 8\,\ca\, \Nbar^{\,-2}
    \,+\, \left[ \left( \fr{73}{2} - 4\,\z2 \right) \ca
          - \fr{7}{3}\:\nf \right]\, \Nbar^{\,0}
    \; ,
\nn \\[1mm]
  c_{\phi, \rm g}^{\,(1,1)} &\!\cong\!&
      -\, 16\,\ca\, \Nbar^{\,-3}
    \,+\, 6\,\z2\,\ca\, \Nbar^{\,-1}
    \; ,
\nn \\[1mm]
  c_{\phi, \rm g}^{\,(2,0)} &\!\cong\!&
          160\:\cas\, \Nbar^{\,-4}
    \,-\, \left[ \fr{440}{3}\: \cas + \fr{16}{3}\: \ca\nf  
          - \fr{32}{3}\: \cf\nf \right]\,\Nbar^{\,-3}
\nn \\[1mm] & & \mbox{}
    \,-\, \left[ \left( \fr{2092}{9} + 16\,\z2 \right) \cas
          - \fr{260}{9}\:\ca\nf - \fr{152}{9}\:\cf\nf \,\right]\, 
            \Nbar^{\,-2}
    \; . \qquad
\label{cPg12}
\eea
The coefficient functions for $F_L$ are, to the lesser accuracy 
required in the present context,
\bea
  c_{L, \rm q}^{\,(1,0)} &\!=\!& 2\,\cf + {\cal O}(\Nbar) \;\; , \quad
  c_{L, \rm q}^{\,(1,1)} \:\:=\:\: {\cal O}(\Nbar^{\,0}) \;\; , \qquad
\nn \\[1mm]
  c_{L, \rm q}^{\,(2,0)} &\!\cong\!&
      -\, \fr{32}{3}\: \cf\nf\, \Nbar^{\,-2}
    \,-\, 8\, \cf\nf\, \Nbar^{\,-1}
\label{cLq12}
\eea
and
\bea
  c_{L, \rm g}^{\,(1,0)} &\!\cong\!&
       \, 4\,\cf\, \Nbar^{\,-1}
    \,-\, 4\,\cf\, \Nbar^{\,0}
    \;\; ,
  \quad c_{L, \rm g}^{\,(1,1)} \;\cong\;
       \, 8\,\cf\, \Nbar^{\,-2}
    \,+\, 8\,\cf\, \Nbar^{\,-1}
    \; ,
\nn \\[1mm]
  c_{L, \rm g}^{\,(2,0)} &\!\cong\!&
      -\, 64\,\cf\ca\, \Nbar^{\,-3}
    \,+\, \left[ \fr{176}{3}\: \cf\ca - 16\,\cfs \,\right] \, \Nbar^{\,-2}
    \; . \qquad
\label{cLg12}
\eea
Note that our normalizations of $c_{T, \rm g}^{}$ and $c_{L, \rm g}^{}$ differ 
by a factor of $1/2$ from those in Refs.~\mbox{\cite{RvN2loop,MM06}}.
Eqs.~(\ref{cTq12}) -- (\ref{cLg12}), and some contributions with a higher 
$\ell+k$ used to further overconstrain the systems of equations discussed 
below Eq.~(\ref{xtoep}), have been obtained from the full $x$-space expressions
in terms of harmonic polylogarithms (HPLs) as discussed in Ref.~\cite{HPLs} and
coded in the {\sc Harmpol} package for {\sc Form} \cite{Form3} together with 
Eq.~(\ref{Mlog}).
 
The corresponding expressions for the NLO and NNLO splitting functions can be
read off directly from Eqs.~(13) and (14) in Ref.~\cite{MV2} and 
Eqs.~(20) -- (23) in Ref.~\cite{AMV1}. For completeness we finally give the
small-$\Nbar$ expansions of the LO splitting functions which we need to 
order $\Nbar^{\,1}$,
\bea
\label{PLOexp}
  P_{\rm qq}^{(0)T} &\!=\!& \left( \fr{5}{2} - 4\,\z2 \right) \cf\, \Nbar
    \,+\, {\cal O}(\Nbar^{\,2})
  \;\; ,\quad
  P_{\rm qg}^{(0)} \:\:=\:\: \fr{4}{3}\:\nf \,-\, \fr{13}{9}\:\nf\, \Nbar
    \,+\, {\cal O}(\Nbar^{\,2}) \; ,
\nn \\[1mm]
  P_{\rm gq}^{(0)T} &\!=\!& 4\,\cf\, \Nbar^{\,-1} \,-\, 3\,\cf 
    \,+\, \fr{7}{2}\:\cf\, \Nbar \,+\, {\cal O}(\Nbar^{\,2}) \; ,
\nn \\[1mm]
  P_{\rm gg}^{(0)T} &\!=\!& 4\,\ca\, \Nbar^{\,-1} 
    \,-\, \fr{11}{3}\:\ca - \fr{2}{3}\:\nf 
    \,+\, \left( \fr{67}{9}\:\ca - 4\,\z2 \right) \ca\, \Nbar 
    \,+\, {\cal O}(\Nbar^{\,2}) \:\: .
\eea
An easy way to obtain the coefficients of any desired positive power of 
$\Nbar$ is to transform the functions to $N$-space harmonic sums \cite{Hsums},
multiply by a sufficiently large power of $\,\Nbar^{\,-1}$, transform back to
$x$-space and proceed as above. Routines for the Mellin transform of the HPLs
and its inverse are also provided by the {\sc Harmpol} package.

Inserting the $N$-space small-$x$ expansions (\ref{cTq12}) -- (\ref{PLOexp}) 
into Eqs.~(\ref{TaCZ}) -- (\ref{ZofP}), we obtain the highest three (two) 
logarithms for the $\as^{\,n}\, \ep^{\,-n+\ell}$, $ \ell = 0,\,1,\, 2\,$
($\ell = 1,\, 2$), contributions to $\widehat{F}_{T, k}$ and 
$\widehat{F}_{\phi, k}$ ($\widehat{F}_{L, k}$) to all orders in $\as$ for which
the higher-order extension of Eq.~(\ref{ZofP}) has been coded.
It turns out that the $\ar^{\,n}$ contributions to $\widehat{F}_{a, \rm g}$ for
$a = T,\,\phi$ can be written as
\beq
\label{FagD}
  \widehat{F}_{a, \rm g}^{\,(n)}(N,\ep) \;\;=\;\; 
  \FR{1}{\ep^{\,2\,n-1}}\; \sum_{\ell=0}^{n-1} 
  \:\: \FR{1}{ N\!-\! 1 - 2\:\!(n-\ell)\:\!\ep }
  \, \big( \, A_{a, \rm g}^{\,(\ell,n)} \,+\, \ep\, B_{a, \rm g}^{\,(\ell,n)} 
  \,+\, \eps\, C_{a, \rm g}^{\,(\ell,n)} \,+\:\ldots\, \big)
\eeq 
or
\beq
\label{FagDx}
  \widehat{F}_{a, \rm g}^{\,(n)}(x,\ep) \;\;=\;\;
  \FR{1}{\ep^{\,2\,n-1}}\; \sum_{\ell=0}^{n-1}
  \:\: x^{\,-1-2\:\!(n-\ell)\:\!\ep}
  \, \big( \, A_{a, \rm g}^{\,(\ell,n)} \,+\, \ep\, B_{a, \rm g}^{\,(\ell,n)}
  \,+\, \eps\, C_{a, \rm g}^{\,(\ell,n)} \,+\:\ldots\, \big) \qquad 
\eeq
up to terms of order $(N-1)^{\,0}$, i.e., non-$x^{\,-1}$ contributions.
Eqs.~(\ref{FagD}) and (\ref{FagDx}) and the corresponding results for 
$\widehat{F}_{T, \rm q}$, $\widehat{F}_{\phi, \rm q}$ and $\widehat{F}_{L,i}$ 
given below form the crucial observation of this article. 
 
Focusing for a moment on the leading logarithms, Eq.~(\ref{FagDx}) decomposes 
$\widehat{F}_{a, \rm g}^{\,(n)}$, which includes terms of the form 
$x^{\,-1} \ln^{\,n+\delta-1\!} x\,$ at all orders $\ep^{\,-n+\delta}$ with 
$\delta = 0,\,1,\,2,\:\ldots\,$, into $\,n$ contributions of the form
\beq
\label{xtoep}
  \ep^{\,-2n+1}\, x^{\,-1-k\,\ep} \;\;=\;\; \ep^{\,-2n+1}\, x^{\,-1}
  \left[ 1 \,-\, k\,\ep \ln x \,+\, \fr12 (k\,\ep)^2 \ln^{\,2\!} x \,+\: 
  \ldots \,\right]
\eeq
with $k = 2,\,4,\, \ldots,\, 2n$. Since $\widehat{F}_{a, \rm g}^{\,(n)}$ only
starts at order $\ep^{\,-n}$, the coefficients $A_{a, \rm g}^{\,(\ell,n)}$ in 
Eq.~(\ref{FagDx}) have to be such that the coefficients of $\,\ep^{\,0},\, 
\ldots\,,\, \ep^{\,n-2}$ in the square bracket in Eq.~(\ref{xtoep}) cancel in 
the sum of these $n$ contributions. Together with the three non-vanishing 
coefficients coefficients of $\ep^{\,-n+\ell}$, $ \ell = 0,\,1,\,2\,$, in 
$\widehat{F}_{a, \rm g}^{\,(n)}$ known from the above NNLO results, we thus 
have an overconstrained system of $n+2$ linear equations for the $n$ 
coefficients $A_{a, \rm g}^{\,(\ell,n)}$ at each order $n$ of the strong 
coupling. 
It is non-trivial that all these systems have solutions, e.g., there would be 
no solutions if the factor of two in front of $(n-\ell)$ in Eqs.~(\ref{FagD}) 
and (\ref{FagDx}) was absent, or if the sign of this term was different.

\pagebreak

The situation is completely analogous for the second and third logarithms. 
The splitting functions and coefficient functions up to NNLO lead to $n+1$
equations for the coefficients $B_{\ell,n\,}$, and to $\,n$ equations for the
coefficients $C_{\ell,n}$ in Eqs.~(\ref{FagD}) and (\ref{FagDx}). Also the
latter system can be overconstrained at all orders except for $n=3$ and $n=4$, 
from which the corresponding contributions to the N$^3$LO coefficient functions 
$c_{a,i}^{\,(3)} \equiv c_{a,i}^{\,(3,0)}$ and splitting functions 
$P^{\,(3)T}_{\!ji}$ are determined.

The decomposition corresponding to Eq.~(\ref{FagD}) for 
$\widehat{F}_{a,\,\rm q}^{\,(n)}$, $a = T,\,\phi$, which are suppressed by one
power of $(N-1)^{\,-1}$ or $\,\ln x\,$ relative to the gluonic quantities,
is given by
\beq
\label{FaqD}
  \widehat{F}_{a, \,\rm q}^{\,(n)}(N,\ep) \;\;=\;\;
  \FR{1}{\ep^{\,2\,n-2}}\; \sum_{\ell=0}^{n-2}
  \:\: \FR{1}{ N\!-\! 1 - 2\:\!(n-\ell)\:\!\ep }
  \, \big( \, A_{a,\,\rm q}^{\,(\ell,n)} \,+\, \ep\, B_{a,\,\rm q}^{\,(\ell,n)}
  \,+\, \eps\, C_{a,\,\rm q}^{\,(\ell,n)} \,+\:\ldots\, \big)
\eeq
for $n > 1$ (there are no $x^{\,-1}$ terms at order $\as$ in these cases, see 
Eqs.~(\ref{cTq12}), (\ref{cPq12}) and (\ref{PLOexp}) above).
The missing equation, due to the lack of an $\ep^{\,-2n+1}$ contribution, is
compensated by the absence of an $x^{\,-1-2\:\!\ep}$ term in the decomposition.
Consequently also the three coefficients written out in Eq.~(\ref{FaqD}) can
be determined from the NNLO quantities given above. 

We have solved the systems of equations for these coefficients and their 
gluonic counterparts in Eq.~(\ref{FagD}) at all orders evaluated for $Z^{\,T}$ in Eq.~(\ref{ZofP}), i.e., to order $\as^{\,16}$.
Re-inserting the results into these equations then determines the respective
highest three logarithms in $\widehat{F}_{a, k}^{\,(n \leq 16)}$ for 
$\,a = T,\,\phi\,$ and $\,k = \rm q,\,g\,$ to all orders in $\ep$, after which
the mass-factorization can be performed to this order in $\as$. 
It is worthwhile to recall that, since the coefficients of 
$\ep^{\,-n},\,\ldots\,,\,\ep^{\,-2}$ at order $\as^{\,n}$ are given in terms of
lower-lower quantities, this process includes a very large number of automatic
checks.  Also these steps have been carried out using {\sc Form} and, for the
more involved last step, {\sc TForm}.
The resulting splitting functions and coefficient functions are presented in
the next two sections. 

Analogous to Eqs.~(\ref{FagD}) and (\ref{FaqD}) the unfactorized partonic 
longitudinal fragmentation functions at all orders $n \geq 2$ can be decomposed
as
\bea
\label{FLgD}
  \widehat{F}_{L, \,\rm g}^{\,(n)}(N,\ep) &\!=\!&
  \FR{1}{\ep^{\,2\,n-2}}\; \sum_{\ell=0}^{n-1}
  \:\: \FR{1}{ N\!-\! 1 - 2\:\!(n-\ell)\:\!\ep }
  \, \big( \, A_{L,\,\rm g}^{\,(\ell,n)} \,+\, \ep\, B_{L,\,\rm g}^{\,(\ell,n)}
  \,+\: \ldots\, \big) \;\; ,
\\[1mm]
\label{FLqD}
  \widehat{F}_{L, \,\rm q}^{\,(n)}(N,\ep) &\!=\!&
  \FR{1}{\ep^{\,2\,n-3}}\; \sum_{\ell=0}^{n-2}
  \:\: \FR{1}{ N\!-\! 1 - 2\:\!(n-\ell)\:\!\ep }
  \, \big( \, A_{L,\,\rm q}^{\,(\ell,n)} \,+\, \ep\, B_{L,\,\rm q}^{\,(\ell,n)}
  \,+\: \ldots\, \big) 
\eea
up to terms of order $(N-1)^{\,0}$. Due to the additional factor of $\ep$ 
relative to the previous cases, the determination of the third coefficients
$\,C_{L,\, i}^{\,(\ell,n)}$ would require the presently unknown third-order 
coefficient functions. The determination of the two highest logarithms in 
$c_{L, \rm i}^{\,(n,0)}$ is performed in the
manner discussed in the previous paragraph, and provides additional checks of
the splitting functions determined from $F_T$ and $F_L$. The resulting
coefficient functions are presented in Section~5.

Like their counterparts for the large-$x$ limit in DIS in Ref.~\cite{ASV1} 
(the publication of the corresponding analysis of SIA is in preparation 
\cite{ALPV1}), see also Refs.~\cite{MvNsoft,MV4}, the decompositions 
(\ref{FagD}) -- (\ref{FLqD}) are inspired by and related (but not identical) to
the decomposition into purely real-emission and the various mixed real-virtual 
contributions. The cancellations of, e.g., the $\ep^{\,-2n+1},\,\ldots\,,
\,\ep^{\,-n+1}$ terms between the $n$ contributions to Eq.~(\ref{FagD}) are
thus related to the KLN theorem \cite{KLN}.

\setcounter{equation}{0}
\section{Resummed timelike splitting functions}
We are now ready to present our (mostly) new all-order small-$x$ results. 
With the exception of graphical illustrations, we will continue to work in 
Mellin-$N$ space. Recall that the connection to \mbox{$x$-space} is simple 
except for the coefficients of $(N-1)^k$ with $k \geq 0$ in the expansion of 
the lowest order quantities about $N=1$ which are required for the all-order
mass factorization. These coefficients are not included in the all-order 
formulae below. 

In this section we present the resummed timelike splitting functions to 
next-to-next-to-leading logarithmic (NNLL) accuracy,
\beq
\label{Pggexp}
  P_{\,\rm ij}^{\,T}(N) \;\:=\;\: \sum_{n=0}^{\infty}\, \ar^{\,n+1} 
  \left(  \delta_{\rm \,ig\,}^{} P_{\,\rm ij,\,LL}^{\,(n)T}(N)
   \:+\: P_{\,\rm ij,\,NLL}^{\,(n)T}(N)
   \:+\: P_{\,\rm ij,\,NNL}^{\,(n)T}(N) 
   \:+\; \ldots \right) \; .
\eeq
The leading log (LL) and next-to-leading log (NLL) 
contributions for $P_{\rm gg}^{\,T}$ and $P_{\rm gq}^{\,T}$ have the form
\beq
\label{PggLL}
   P_{\rm gg,\,LL}^{\,(n)T}(N) \;\:=\;\: 
   \frac{\ca}{\cf}\: P_{\rm gq,\,LL}^{\,(n)T}(N) \;\:=\;\:
   - \; \frac{(-8\,C_A)^{n+1}}{2(N-1)^{2n+1}}\: A^{(n)}_{\rm gi} 
\eeq
and
\bea
\label{PggNL}
   P_{\rm gg,\,NLL}^{\,(n)T}(N) &\!=\!&
   - \; \frac{(-8)^{\,n}\,C_A^{\,n-1}}{3(N-1)^{2n}} \:
   \Big[ (11\,\cas + 2\,\ca \nf ) \, B^{\,(n)}_{\rm gg,1}
   \,-\, 2\: \cf \nf\, B^{\,(n)}_{\rm gg,2} \Big] \; ,
\\[1mm]
\label{PgqNL}
   P_{\rm gq,\,NLL}^{\,(n)T}(N) &\!=\!&
   -\; \frac{(-8)^{n}\,C_A^{\,n-2}\,\cf}{3(N-1)^{2n}} \:
   \Big[ \cas\, B^{\,(n)}_{\rm gq,1} 
   \,+\, 2\,\ca \nf\, B^{\,(n)}_{\rm gq,2}
   \,-\, 2\,\cf \nf\, B^{\,(n)}_{\rm gq,3} \Big] \; .
\eea
The coefficients in Eqs.~(\ref{PggLL}) -- (\ref{PgqNL}) have been determined
to order $\as^{\,16}$ ($n=15$ in Eq.~(\ref{Pggexp})), and are given in Table 
\ref{tab:pginl} to the tenth order in $\as$ -- for the next six orders see 
the text below Eq.~(\ref{Aqifrom10}).
The the highest two contributions to $P_{\rm qg}^{\,T}$ and $P_{\rm qq}^{\,T}$ 
can be written as
\beq
\label{PqgLL}
   P_{\rm qg,\,NLL}^{\,(n)T}(N) \;\:=\;\:
   \frac{\ca}{\cf}\: P_{\rm qq,\,NLL}^{\,(n)T}(N) \;\:=\;\:
   \frac{(-8\,C_A)^n \,\nf}{3(N-1)^{2n}}\; 2\,A^{\,(n)}_{\rm qi}
\eeq
and
\bea
\label{PqgNL}
   P_{\rm qg,\,NNL}^{\,(n)T}(N) &\!=\!& - \;
   \frac{(-8)^{n}\,C_A^{\,n-2}\,\nf}{9(N-1)^{2n-1}} \:
   \Big[ \cas\, B^{\,(n)}_{\rm qg,1}
   \,+\, \ca \nf\, B^{\,(n)}_{\rm qg,2}
   \,-\, \cf \nf\, B^{\,(n)}_{\rm qg,3} \Big] \; ,
\\[1mm]
\label{PqqNL}
   P_{\rm qq,\,NNL}^{\,(n)T}(N) &\!=\!& - \;
   \frac{(-8)^{n}\,C_A^{\,n-3}\,\cf \nf}{9(N-1)^{2n-1}} \:
   \Big[ \cas\, B^{\,(n)}_{\rm qq,1}
   \,+\, \ca \nf\, B^{\,(n)}_{\rm qq,2}
   \,-\, \cf \nf\, B^{\,(n)}_{\rm qq,3} \Big] \; .
\eea
The coefficients in Eqs.~(\ref{PqgLL}) -- (\ref{PqqNL}) are given in Table 
\ref{tab:pqinl} to the sixteenth order in $\as$, for brevity using a numerical
form for $n \geq 12$.
\begin{table}[thb]
\small
\vspace*{1mm}
\begin{center}
\begin{tabular}{|c|rrc|rrc|c|}\hline
 & & & & & & & \\[-3mm]
 $n$ &
 $A^{\,(n)}_{\rm gi}$   & $B^{\,(n)}_{\rm gg,1}$ &
 $B^{\,(n)}_{\rm gg,2}$ & $B^{\,(n)}_{\rm gq,1}$ & 
 $B^{\,(n)}_{\rm gq,2}$ & $B^{\,(n)}_{\rm gq,3}$ & $A^{\,(n)}_{\rm qi}$ \\[1mm] \hline
 & & & & & & & \\[-2.5mm]
$0$ & 1      & 1       & --                      & 9        & --     & --                      & -- \\[1mm]
$1$ & 1      & 1       & 2                       & 9        & --     & --                      & -- \\[1mm]
$2$ & 2      & 3       & 5                       & 29       & 1      & 1                       & 1 \\[1mm]
$3$ & 5      & 10      & $\FR{49}{3}$            & 100      & 5      & $\FR{19}{3}$            & $\FR{11}{3}$ \\[3.5mm]
$4$ & 14     & 35      & $\FR{347}{6}$           & 357      & 21     & $\FR{179}{6}$           & $\FR{73}{6}$ \\[3.5mm]
$5$ & 42     & 126     & $\FR{6353}{30}$         & 1302     & 84     & $\FR{3833}{30}$         & $\FR{1207}{30}$ \\[3.5mm]
$6$ & 132    & 462     & $\FR{11839}{15}$        & 4818     & 330    & $\FR{7879}{15}$         & $\FR{2021}{15}$ \\[3.5mm]
$7$ & 429    & 1716    & $\FR{624557}{210}$      & 18018    & 1287   & $\FR{444377}{210}$      & $\FR{96163}{210}$ \\[3.5mm]
$8$ & 1430   & 6435    & $\FR{316175}{28}$       & 67925    & 5005   & $\FR{236095}{28}$       & $\FR{44185}{28}$ \\[3.5mm]
$9$ & 4862   & 24310   & $\FR{54324719}{1260}$   & 257686   & 19448  & $\FR{42072479}{1260}$   & $\FR{6936481}{1260}$  \\[3.5mm]
\hline
\end{tabular}
\vspace{1.5mm}
\caption{The coefficients of the LL and NLL small-$x$ approximations in 
$N$-space (\ref{PggLL}) -- (\ref{PgqNL}) for the timelike gluon-gluon and
gluon-quark splitting functions for the first ten orders in $\as$. Also shown
(last column) are the related NLL quark-parton coefficients in 
Eq.~(\ref{PqgLL}).
\label{tab:pginl}}
\end{center}
\vspace*{-4mm}
\end{table}

\begin{table}[thb]
\footnotesize
\vspace*{1mm}
\begin{center}
\begin{tabular}{|r|ccc|ccc|}\hline
 & & & & & & \\[-3mm]
 $n$ &
 $B^{\,(n)1}_{\rm qg}$ & $B^{\,(n)2}_{\rm qg}$ & $B^{\,(n)3}_{\rm qg}$ &
 $B^{\,(n)1}_{\rm qq}$ & $B^{\,(n)2}_{\rm qq}$ & $B^{\,(n)3}_{\rm qq}$ \\[2mm] \hline
 & & & & & & \\[-2mm]
 2  & 2                         & 1                         & 2                          & 1                           & --                         & -- \\[1.5mm]
 3  & $\FR{89}{3}$              & $\FR{17}{3}$              & 10                         & 26                          & 2                          & $\FR{8}{3}$ \\[3.5mm]
 4  & $\FR{863}{6}$             & $\FR{49}{2}$              & $\FR{253}{6}$              & $\FR{395}{3}$               & $\FR{37}{3}$               & $\FR{107}{6}$ \\[3.5mm]
 5  & $\FR{3028}{5}$            & $\FR{997}{10}$            & $\FR{5153}{30}$            & $\FR{16961}{30}$            & $\FR{892}{15}$             & $\FR{913}{10}$ \\[3.5mm]
 6  & $\FR{219389}{90}$         & $\FR{3574}{9}$            & $\FR{20701}{30}$           & $\FR{207263}{90}$           & $\FR{11807}{45}$           & $\FR{12617}{30}$ \\[3.5mm]
 7  & $\FR{6069467}{630} $      & $\FR{197447}{126}$        & $\FR{57770}{21}$           & $\FR{412927}{45}$           & $\FR{349373}{315}$         & $\FR{64299}{35}$ \\[3.5mm]
 8  & $\FR{38066605}{1008}$     & $\FR{2216267}{360}$       & $\FR{9169289}{840}$        & $\FR{36475945}{1008}$       & $\FR{11537219}{2520}$      & $\FR{6518189}{840}$ \\[3.5mm]
 9  & $\FR{35395649}{240}$      & $\FR{8688247}{360}$       & $\FR{7255001}{168}$        & $\FR{143112541}{1008}$      & $\FR{46944767}{2520}$      & $\FR{81079091}{2520}$ \\[3.5mm]
 10 & $\FR{1449188057}{2520}$   & $\FR{595482761}{6300}$    & $\FR{537123949}{3150}$     & $\FR{22226523}{40}$         & $\FR{157729997}{2100}$     & $\FR{59261597}{450}$ \\[3.5mm]
 11 & $\FR{41422167289}{18480}$ & $\FR{17097960349}{46200}$ & $\FR{93203672711}{138600}$ & $\FR{120436395671}{55440}$  & $\FR{41718615557}{138600}$ & $\FR{8228126859}{15400}$ \\[3.5mm]
 12 & {\small 8.7381596\,10$^6\!$} & {\small 1.4491517\,10$^6\!$} & {\small 2.6500373\,10$^6\!$} 
    & {\small 8.4902823\,10$^6\!$} & {\small 1.2012745\,10$^6\!$} & {\small 2.1542828\,10$^6\!$} \\[2mm]
 13 & {\small 3.4082509\,10$^7\!$} & {\small 5.6761999\,10$^6\!$} & {\small 1.0438829\,10$^7\!$} 
    & {\small 3.3186784\,10$^7\!$} & {\small 4.7804744\,10$^6\!$} & {\small 8.6473779\,10$^6\!$} \\[2mm]
 14 & {\small 1.3302661\,10$^8\!$} & {\small 2.2242725\,10$^7\!$} & {\small 4.1111839\,10$^7\!$} 
    & {\small 1.2976945\,10$^8\!$} & {\small 1.8985567\,10$^7\!$} & {\small 3.4597524\,10$^7\!$} \\[2mm]
 15 & {\small 5.1960779\,10$^8\!$} & {\small 8.7203656\,10$^7\!$} & {\small 1.6190693\,10$^8\!$} 
    & {\small 5.0769752\,10$^8\!$} & {\small 7.5293381\,10$^7\!$} & {\small 1.3808638\,10$^8\!$} \\[1.5mm]
\hline
\end{tabular}
\vspace{-1mm}
\caption{The corresponding coefficients in Eqs.~(\ref{PqgLL}) -- (\ref{PqqNL})
for the timelike quark-gluon and quark-quark splitting functions to the 
sixteenth order in the strong coupling constant.
\label{tab:pqinl}}
\vspace*{-4mm}
\end{center}
\end{table}

The general form and generating function for these series are known at this 
point (to this author) only for Eq.~(\ref{PggLL}) and the non-$C_F$ terms in 
the square brackets in Eqs.~(\ref{PggNL}) and (\ref{PgqNL}), i.e., those 
entries that do not involve factorial denominators. 
$A^{(n)}_{\rm gi}$ are the Catalan numbers \cite{OEIS,Catalan},
\beq
\label{Agi}
  A^{(n)}_{\rm gi} \;=\; \frac{ (2n)! }{ n! (n+1)! } \;=\; {1 \over n+1}
  \left( \begin{array}{cc} \!\! 2\:\!n \!\! \\ n \end{array} \right) \; .
\eeq
$B^{\,(n)}_{\rm gg,1}$ and $B^{\,(n)}_{\rm gq,2}$ are given by 
\cite{binom12}
\beq
\label{Bgg1}
  B^{\,(n)}_{\rm gg,1} \;=\; \left( \begin{array}{cc} \!\! 2\:\!n-1 \!\!
                             \\ n \end{array} \right) 
\;, \quad
  B^{\,(n)}_{\rm gq,2} \;=\; \left( \begin{array}{cc} \!\! 2\:\!n-1 \!\! 
                             \\ n-2 \end{array} \right) 
  \;=\; B^{\,(n)}_{\rm gg,1} \,-\, A^{(n)}_{\rm gi} \; ,
\eeq
and the remaining coefficient in Eq.~(\ref{PgqNL}) is related to these results
by
\beq
\label{Bgq1}
  B^{\,(n)}_{\rm gq,1} \;\:=\;\: 11\,B^{\,(n)}_{\rm gg,1} - 2\,A^{(n)}_{\rm gi} 
\; .
\eeq
Furthermore it is interesting to note that the last entries in 
Eqs.~(\ref{PggNL}) and (\ref{PgqNL}) have a much simpler difference,
\beq
\label{Bgq2qg3}
  B^{\,(n)}_{\rm gg,2} \,-\, B^{\,(n)}_{\rm gq,3}\;\:=\;\: 2\, A^{(n)}_{\rm gi} 
 \; ,
\eeq
and that the quark-parton coefficients in Eq.~(\ref{PqgLL}) are related to the 
above quantities by 
\beq 
\label{AqitoBgg}
  A^{(n)}_{\rm qi} + B^{\,(n)}_{\rm gg,2} \;\:=\;\: 2\,B^{\,(n)}_{\rm gg,1} \; .
\eeq
Hence only one more complicated series is contained in $B^{\,(n)}_{\rm gg,2\,}$,
$B^{\,(n)}_{\rm gq,3}$ and $A^{(n)}_{\rm qi}$; and an analytic formula for
any of these quantities would lead to closed expressions for all 
$\as^{\,n+1\!}/ (N-1)^{2n-1}$ contributions to the timelike splitting functions.
The coefficients in Eq.~(\ref{PqgLL}) for $n=10, \ldots,15$ are
\small
\bea
\label{Aqifrom10}
  A^{(10)}_{\rm qi} &\!\!=\!\!& \frac{12229277}{630} \;\; , \quad\quad\;\;
  A^{(11)}_{\rm qi} \;\:=\;\: \frac{136789507}{1980} \;\; , \quad\;\;
  A^{(12)}_{\rm qi} \;\,=\; \frac{245398487}{990} \;\; , \quad
\nn \\[1mm]
  A^{(13)}_{\rm qi} &\!\!=\!\!& \frac{16139182231}{18018} \;\; , \quad
  A^{(14)}_{\rm qi} \;\:=\;\: \frac{6986603759}{2145} \;\; , \quad
  A^{(15)}_{\rm qi} \;=\;\: \frac{102190158383}{8580} \;\; .
\eea
\normalsize
The corresponding expressions for Eqs.~(\ref{PggLL}) -- (\ref{PgqNL}) can 
be inferred from Eqs.~(\ref{Agi}) -- (\ref{AqitoBgg}).

The results (\ref{Agi}) -- (\ref{Bgq1}) lead to the closed NLL expressions
\bea
\label{Pgg-cl}
  P_{\,\rm gg}^{\,T}(N) \Big|_{C_F=0} \!&\!=\!&
  \left\{ \! \left( 1 - 4\, \xi \right)^{1/2}\! - 1 \right\} 
  \; \fr{1}{4} \: (N-1)
\nn \\[1mm] & & \mbox{\hspn} - \: 
  \left\{ \left( 1 -4\,\xi \right)^{-1/2}\! + 1 \right\}
  \,\ar \, \Big( \,\fr{11}{6}\:\ca + \fr{1}{3}\:\nf \Big)   
  \;+\; P_{\,\rm gg,\,NNL}^{\,(n)T}(N) \; , 
\eea
and
\bea
\label{Pgq-cl}
 \bigg[ \:\frac{\ca}{\cf}\: P_{\,\rm gq}^{\,T}(N) \bigg]_{C_F=0}^{\rm NLL} 
 \!\!&\!=\!& 
  \left\{ \! \left( 1 - 4\,\xi \right)^{1/2}\! - 1 \right\}
  \, \fr{1}{24} \: (N-1)^2 \, \left( 1 + \nf / \ca \right) \qquad
\nn \\[1mm] & & \mbox{\hspn} 
  \,-\, 
  \left\{ \left( 1 - 4\,\xi \right)^{-1/2}\! + 1 \right\}
  \: \left( \fr{11}{6}\:\ca + \fr{1}{3}\:\nf \right) \,\ar 
\eea
%
%
with
\beq
  \xi \:\;=\:\; -\,\frac{8\:\! \ca \ar}{(N-1)^2}
  \quad \mbox{ and } \quad
  \ar \:\;\equiv\:\; \frac{\as}{4\pi} \;\; .
\eeq
The first line of Eq.~(\ref{Pgg-cl}) and the directly related LL part of 
$P_{\,\rm gq}^{\,T}(N)$ agree, of course, with the classical result 
(\ref{PLLsum}) of 
Refs.~\cite{LLxto0}. Already at order $\as^{\,3}$ \cite{MV2,AMV1}, the NLL
second line of Eq.~(\ref{Pgg-cl}) is not the same as the result in Ref.\
\cite{MuellerNL} which does not refer to the \MSb\ scheme, see
Ref.~\cite{ABKK11}.

The expressions for the third logarithms (NNLL for $P_{\rm gi}$ and 
N$^{\:\!3}$LL for $P_{\rm qi}$) are far more lengthy. 
We therefore confine ourselves here to the full analytic expressions at order 
$\as^{\,4}$, and present the higher-order coefficients only in numerical form 
for the case of QCD, $\ca = 3$ and $\cf = 4/3$. The leading $N \!\ra\! 1$ 
behaviour of $P_{\rm gg}^{\,(3)T}$ and $P_{\rm gq}^{\,(3)T}$ is given by
\bea
  P_{\rm gg}^{\,(3)T}(N) &\!=\!& 
    -\; {512 \over (N-1)^7}\; 20\,\cafo
  \:+\: {512 \over (N-1)^6} \: \Big\{ 
        \fr{110}{3}\: \cafo + \fr{20}{3}\: \cath \nf 
        - \fr{98}{9}\: \cas\cf\nf \Big\}
\nn \\[0.5mm] & & \mbox{} \!\!
  \:+\: {512 \over (N-1)^5} \: \Big\{ \!
        \left( - \fr{899}{18} + 16\,\z2 \right)\, \cafo
        - 15\, \cath \nf + \fr{76}{3}\, \cas\cf\nf - \fr{2}{3}\: \cas \nfs 
\nn \\[-1mm] & & \mbox{\hspace*{2.4cm}}
        + \fr{14}{9}\: \ca\cf\nfs - \fr{16}{27}\:\cfs\nfs
        \Big\} \:\:+\:\: \ldots
\eea
with $\z2 = \pi^2/6$ and
\bea
  P_{\rm gq}^{\,(3)T}(N) &\!=\!&
    -\; {512 \over (N-1)^7}\; 20\,\cath\cf
  \:+\: {512 \over (N-1)^6} \: \Big\{
        \fr{100}{3}\: \cath\cf + \fr{10}{3}\: \cas\cf\nf
        - \fr{38}{9}\: \cas\cf\nf \Big\}
\nn \\[0.5mm] & & \mbox{} \!\!
  \:+\: {512 \over (N-1)^5} \: \Big\{ \!
        \left( - \fr{110}{3} + \fr{26}{3}\,\z2 \right)\, \cath\cf
        - \left( \fr{55}{12} - \fr{22}{3}\,\z2 \right)\, \cas\cfs
        -  \fr{793}{108}\, \cas\cf \nf 
\nn \\[-1mm] & & \mbox{\hspace*{2.4cm}}
        + \fr{221}{27}\, \ca\cfs\nf - \fr{1}{9}\: \ca\cf \nfs
        + \fr{4}{27}\: \cfs\nfs 
        \Big\} \:\:+\:\: \ldots \;\; .
\eea
The corresponding results for $P_{\rm qg}^{\,(3)T}$ and $P_{\rm qq}^{\,(3)T}$ 
read
\bea
  P_{\rm qg}^{\,(3)T}(N) &\!=\!&
    -\; {512 \over (N-1)^6}\; \fr{22}{9}\: \cath\nf
  \:+\: {512 \over (N-1)^5} \: \Big\{
        \fr{89}{27}\: \cath\nf + \fr{17}{27}\: \cas\nfs 
        - \fr{10}{9}\: \ca\cf\nfs \Big\}
\nn \\[0.5mm] & & \mbox{} \!\!
  \:+\: {512 \over (N-1)^4} \: \Big\{ \!
        \left( - \fr{187}{72} + \fr{4}{3}\:\z2 \right)\, \cath\nf
        + \left( \fr{1}{9} + \fr{2}{9}\:\z2 \right)\, \cas\cf\nf
        - \fr{23}{18}\, \cas\nfs 
\nn \\[-1mm] & & \mbox{\hspace*{2.4cm}}
        + \fr{185}{81}\, \ca\cf\nfs 
        - \fr{1}{27}\: \ca\nft + \fr{2}{27}\:\cf\nft
        \Big\} \:\:+\:\: \ldots
\eea
and
\bea
  P_{\rm qq}^{\,(3)T}(N) &\!=\!&
    -\; {512 \over (N-1)^6}\; \fr{22}{9}\: \cas\cf\nf
  \:+\: {512 \over (N-1)^5} \: \Big\{
        \fr{26}{9}\: \cas\cf\nf + \fr{2}{9}\: \ca\cf\nfs
        - \fr{8}{27}\: \cfs\nfs \Big\}
\nn \\[0.5mm] & & \mbox{} \!\!
  \:+\: {512 \over (N-1)^4} \: \Big\{ \!
        \left( - \fr{763}{648} + \fr{4}{9}\:\z2 \right)\, \cas\cf\nf
        - \left( \fr{4}{9} - \fr{10}{9}\:\z2 \right)\, \ca\cfs\nf
\nn \\[-1mm] & & \mbox{\hspace*{2.4cm}}
        - \fr{46}{81}\, \ca\cf\nfs + \fr{50}{81}\:\cfs\nfs
        \Big\} \:\:+\:\: \ldots \;\; ,
\eea
where the dots indicate terms beyond the present accuracy of the expansion in 
powers of $1/(N-1)$. The respective NNLL and N$^{\:\!3}$LL higher-order 
expressions are written as
\bea
\label{PgiNNL}
   P_{\rm gi,\,NNL}^{\,(n)T}(N) &\!=\!&
   {(-1)^n \over (N-1)^{2n-1}}\: \left( 
     96^{\,n}\, C^{\,(n)}_{\rm gi,0} \,+\, 
     96^{\,n-1}\, C^{\,(n)}_{\rm gi,1}\: \nf \,+\,
     96^{\,n-2}\, C^{\,(n)}_{\rm gi,2}\: \nfs 
   \right) 
\eea
and
\bea
\label{PqiNNL}
   P_{\rm qi,\,N^3L}^{\,(n)T}(N) &\!=\!&
   {(-1)^n \over (N-1)^{2n-2}}\: \left(
     96^{\,n-1}\, C^{\,(n)}_{\rm qi,1}\: \nf \,+\,
     96^{\,n-2}\, C^{\,(n)}_{\rm qi,2}\: \nfs \,+\,
     96^{\,n-3}\, C^{\,(n)}_{\rm qi,3}\: \nft
   \right) \; .
\eea
The coefficients for Eq.~(\ref{PgiNNL}) and Eq.~(\ref{PqiNNL}) are given in 
Table 3 and Table 4, respectively. Here the relative normalization of the 
coefficients of different orders in $\as$ is such that the ratios
$C^{\,(n)}_{\rm ij,l}/C^{\,(n-1)}_{\rm ij,l}$ will tend to one for 
$n \ra \infty$, if the NNLL correction have the same convergence properties
as the LL and NLL contributions in Eqs.~(\ref{Pgg-cl}) and (\ref{Pgq-cl}).
The present calculations have not been carried out to an order sufficient
to definitely decide whether this is indeed the case.

\begin{figure}[p]
\vspace*{-1mm}
\centerline{\epsfig{file=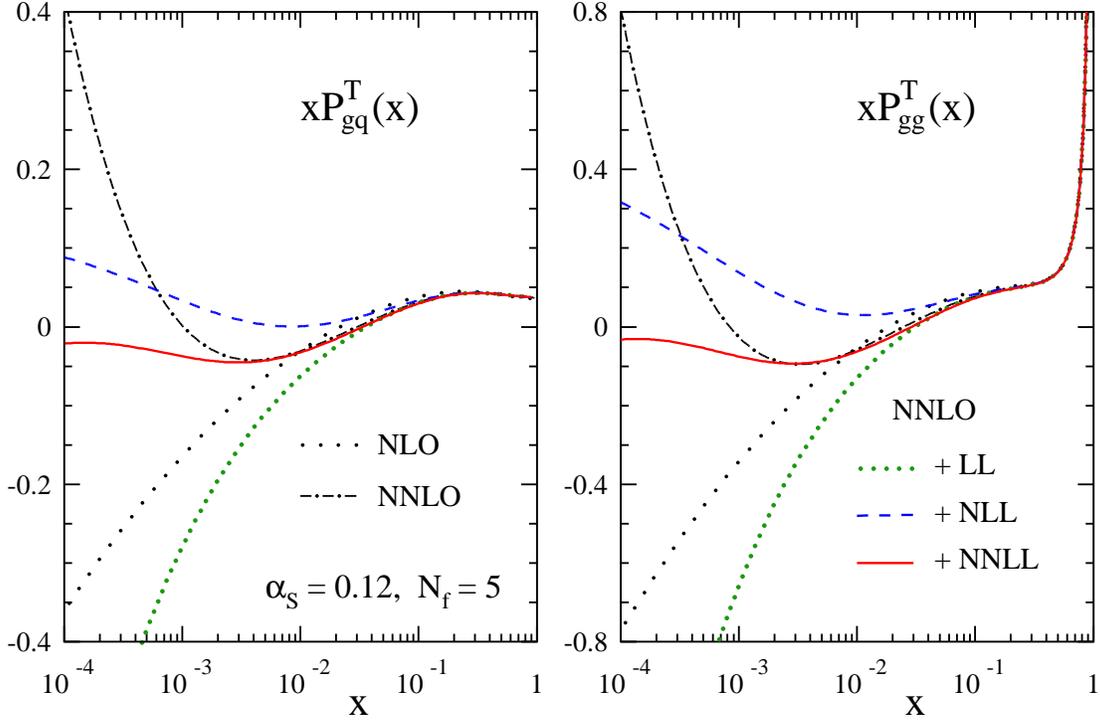,width=14.8cm,angle=0}\quad}
\vspace{-3mm}
\caption{ \label{fig:Pgitexp}
The timelike gluon-quark and gluon-gluon splitting functions at a typical value
of the strong coupling constant $\as$, multiplied by $x$ for display purposes. 
Shown are the NLO and NNLO approximations, and the consequences of adding the
leading ($\as^{\,n-1} \ln^{\,2n\!} x$), next-to-leading and 
next-to-next-to-leading small-$x$ logarithms to the latter at all higher orders
in $\as$.}
\vspace{-2mm}
\end{figure}

\begin{figure}[p]
\vspace{-1mm}
\centerline{\epsfig{file=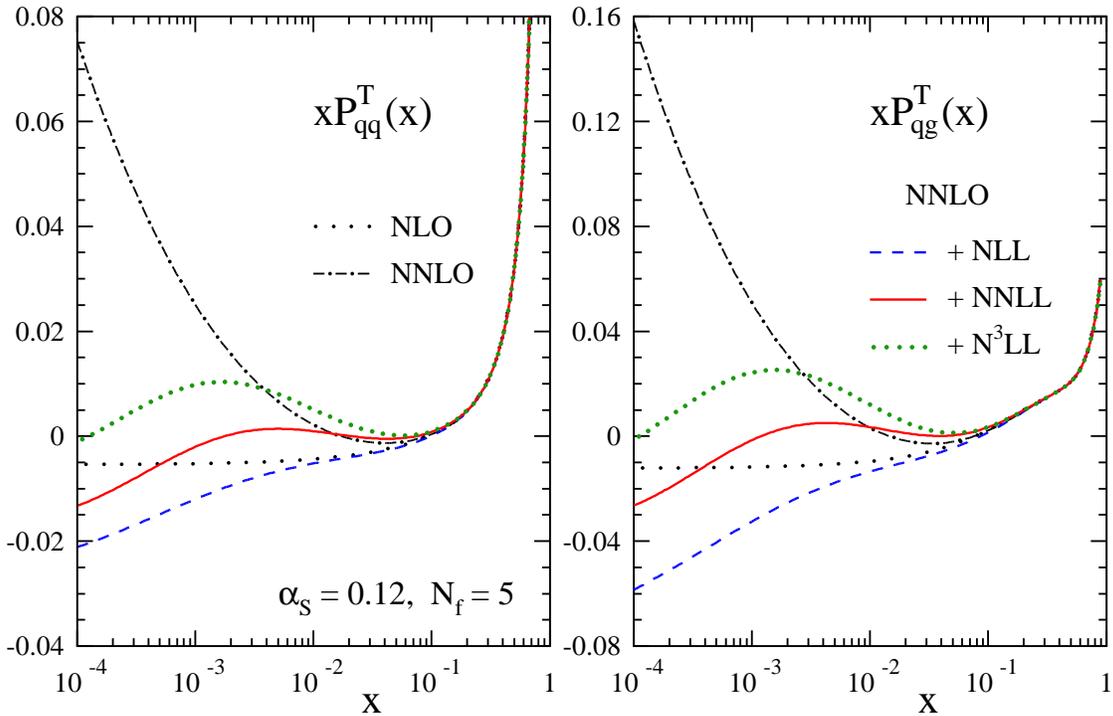,width=14.8cm,angle=0}\quad}
\vspace{-3mm}
\caption{ \label{fig:Pqitexp}
 As Fig.~\ref{fig:Pgitexp}, but for the timelike quark-quark and quark-gluon 
 splitting functions, where the highest logarithms are of the  next-to-leading
 logarithmic form $\as^{\,n-1} \ln^{\,2n-1\!} x$.}
\vspace{-2mm}
\end{figure}

\begin{figure}[p]
\vspace*{-1mm}
\centerline{\epsfig{file=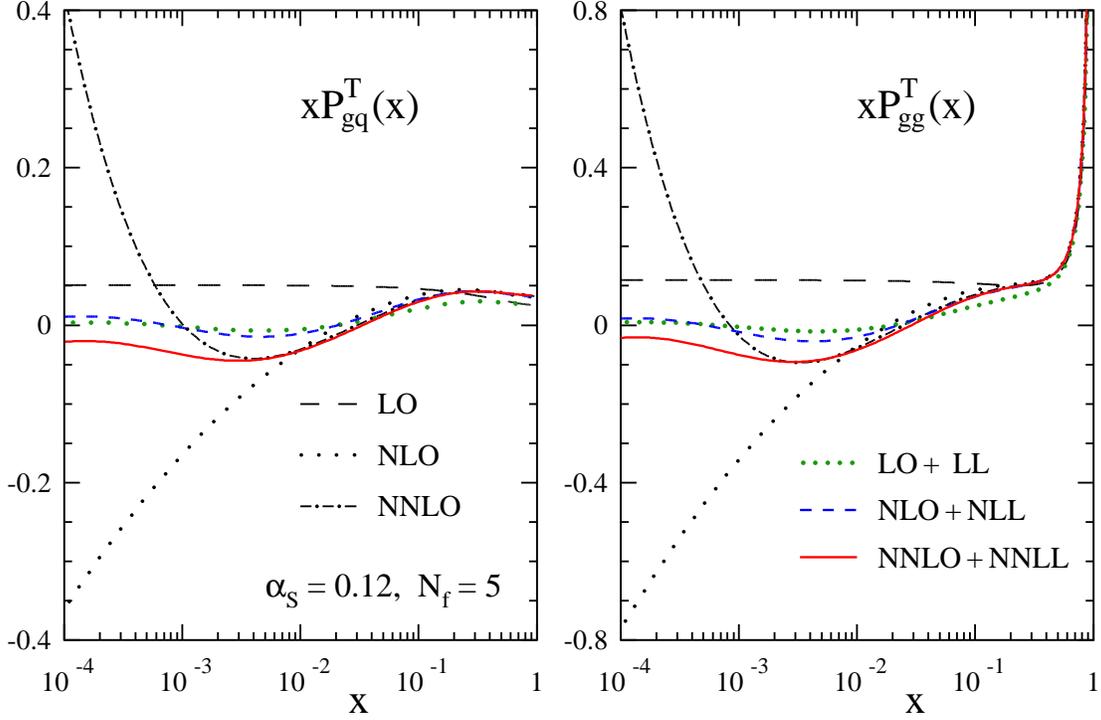,width=14.8cm,angle=0}\quad}
\vspace{-3mm}
\caption{ \label{fig:Pgitexp2}
The timelike gluon-quark and gluon-gluon splitting functions at a typical value
of $\as$, multiplied by $x$ for display purposes.
The LO, NLO and NNLO fixed-order approximations are compared with the small-$x$ 
resummed results obtained by respectively adding the LL, NLL and NNLL 
contributions at all numerically relevant higher orders in $\as$.}
\vspace{-2mm}
\end{figure}

\begin{figure}[p]
\vspace{-1mm}
\centerline{\epsfig{file=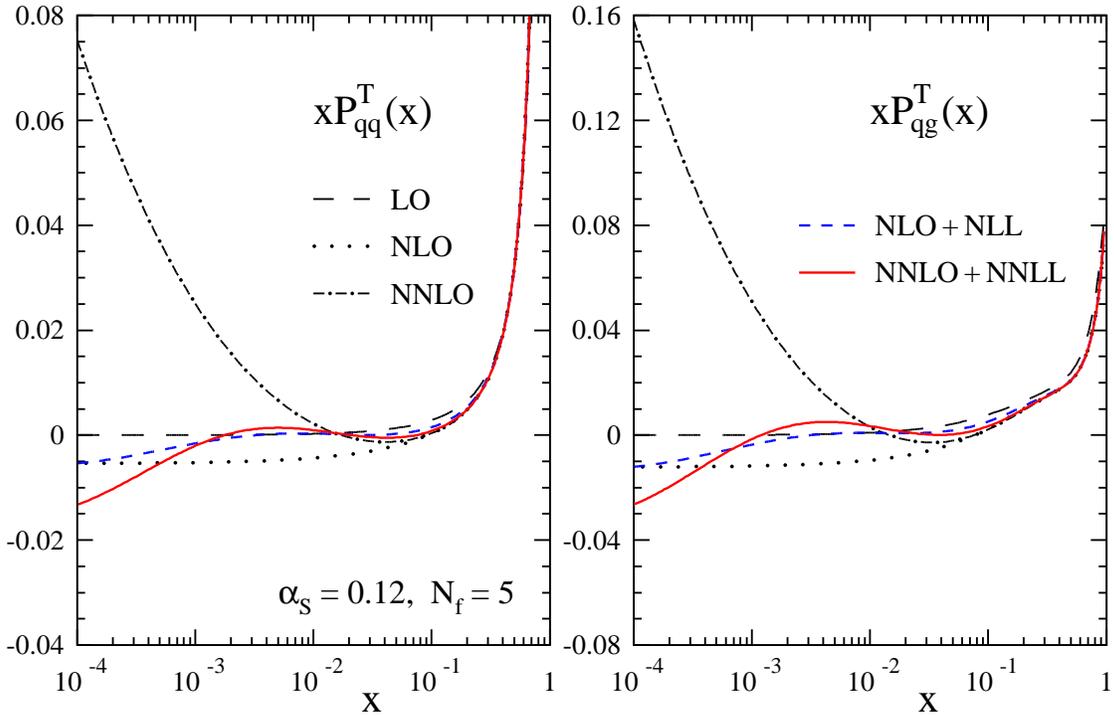,width=14.8cm,angle=0}\quad}
\vspace{-3mm}
\caption{ \label{fig:Pqitexp2}
 As Fig.~\ref{fig:Pgitexp2}, but for the timelike quark-quark and quark-gluon
 splitting functions which do not receive leading logarithmic (LL)
 corrections. Hence only two resummed curves are shown.}
\vspace{-2mm}
\end{figure}

The fixed-order and resummed timelike splitting functions are illustrated and
compared in Figs.~\ref{fig:Pgitexp} -- \ref{fig:Pqitexp2} at a standard
reference scale, $\Qs \simeq M_Z^{\,2\,}$, for $\nf = 5$ effectively massless
flavours. 
For the corresponding value $\as \simeq 0.12$ of the strong coupling constant,
the expansions to order $\as^{\:\!16}$ are sufficient, and for some of the NNLL
results required, for an accuracy of 0.1\% or better down to the lowest 
$x$-values shown, $x = 10^{\,-4}$.
An extension of the maximal order to cover one more order of magnitude in $x$ 
is definitely feasible, but does not appear to be warranted for any foreseeable
analyses of experimental data.

It is clear from Figs.~\ref{fig:Pgitexp} and \ref{fig:Pqitexp} that the 
available fixed-order approximations to the splitting functions are not reliable
at $x \lsim 10^{\,-3}$ for the gluon-parton cases -- recall Eq.~(\ref{Devol})
and the form (\ref{PTmat}) of the timelike splitting function matrix, which is
transposed relative to the spacelike case of the initial-state parton 
distributions -- and $x \lsim 10^{\,-2}$ for the quark-parton cases. 
Obviously it is also insufficient to only add the previously known leading-%
logarithmic resummation \cite{LLxto0} from order $\as^{\:\!4}$ to the NNLO
gluon-quark and gluon-gluon splitting functions in Fig.~\ref{fig:Pgitexp}. 
On the other hand, a near-perfect cancellation of the strong $x$-dependences 
is exhibited by the NNLO$\,+\,$NNLL results for $xP_{\!ji}^{\,T}$ especially in
these cases. 
The situation is somewhat less clear-cut for the quark-parton splitting 
functions in Fig.~\ref{fig:Pqitexp} where, as already at order $\as^{\:\!3}$ 
but unlike the gluon-parton cases, the effects of the second and third 
logarithms have the same sign. 
Within the present uncertainties all results appear to be consistent
with $\,x P_{\!ji}^{\,T} \,\approx\, 0\,$ at $x < 10^{\,-2}$.

In Figs.~\ref{fig:Pgitexp2} and \ref{fig:Pqitexp2} the known three fixed-order
approximations are compared by their resummed counterparts obtained by adding
the `appropriate' higher-order resummations to the respective fixed-order
results, i.e., forming the LO$\,+\,$LL (for the gluon-parton cases), 
NLO$\,+\,$NLL and NNLO $+\,$NNLL combinations. 
The differences between the two expansions at $x < 10^{\,-2}$ are striking. 
Some questions remain due to the relatively large NNLO$\,+\,$NNLL corrections 
in Fig.~\ref{fig:Pgitexp2} and the corresponding behaviour at $x < 10^{\,-3}$ 
in Fig.~\ref{fig:Pqitexp2}. 
Their answer will require the calculation of the fourth-order (N$^3$LO) 
splitting functions (from which the N$^3$LL resummations for $P_{\rm gq}^{\,T}$ 
and $P_{\rm gg}^{\,T}$ can be inferred analogously to the present calculations)
which, unfortunately, is not expected in the near future. 
In the meantime the NNLO$\,+\,$NNLL results, and their comparison with the 
previous NLO$\,+\,$NLL resummed order, should be sufficient for practical data 
analysis including estimates of the effect of the presently unknown higher 
orders.

\begin{table}[p]
\small
\begin{center}
\begin{tabular}{|c|rrr|rrr|}\hline
 & & & & & & \\[-3mm]
 $n$ &
 $C^{\,(n)}_{\rm gg,0}$ & $C^{\,(n)}_{\rm gg,1}$ &
 $C^{\,(n)}_{\rm gg,2}$ & $C^{\,(n)}_{\rm gq,0}$ &
 $C^{\,(n)}_{\rm gq,1}$ & $C^{\,(n)}_{\rm gq,2}$ \\[2mm] \hline
 & & & & & & \\[-2mm]
$1$ & 0         & 3.4074074 & --      &$-$0.3148148 & --        & --        \\[1mm]
$2$ & 0.7923411 & 5.4814815 & 2.3703704 & 0.2174233 & 2.2469136 & --        \\[1mm]
$3$ & 1.1074453 & 5.6111111 & 4.4334705 & 0.3976321 & 2.4698217 & 0.9657064 \\[1mm]
$4$ & 1.3336401 & 5.6790123 & 5.0809328 & 0.5129934 & 2.5064300 & 1.3924707 \\[1mm]
$5$ & 1.5204469 & 5.7204475 & 5.2181070 & 0.6035361 & 2.5046339 & 1.5768328 \\[1mm]
$6$ & 1.6839029 & 5.7522248 & 5.1713306 & 0.6809114 & 2.4969915 & 1.6546258 \\[1mm]
$7$ & 1.8313932 & 5.7823077 & 5.0603175 & 0.7498896 & 2.4920010 & 1.6831461 \\[1mm]
$8$ & 1.9670281 & 5.8140519 & 4.9316765 & 0.8128873 & 2.4915980 & 1.6880472 \\[1mm]
$9$ & 2.0933792 & 5.8486686 & 4.8040710 & 0.8713173 & 2.4957684 & 1.6815451 \\[1mm]
$10$& 2.2121870 & 5.8864117 & 4.6847584 & 0.9260927 & 2.5039523 & 1.6697449 \\[1mm]
$11$& 2.3246982 & 5.9271265 & 4.5761405 & 0.9778468 & 2.5154984 & 1.6558008 \\[1mm]
$12$& 2.4318435 & 5.9705020 & 4.4785239 & 1.0270426 & 2.5298070 & 1.6413704 \\[1mm]
$13$& 2.5343410 & 6.0161860 & 4.3913363 & 1.0740322 & 2.5463666 & 1.6273227 \\[1mm]
$14$& 2.6327593 & 6.0638358 & 4.3136775 & 1.1190916 & 2.5647543 & 1.6140995 \\[1mm]
$15$& 2.7275579 & 6.1131386 & 4.2445696 & 1.1624423 & 2.5846246 & 1.6019061 \\[1mm]
\hline
\end{tabular}
\vspace{1.0mm}
\caption{The numerical coefficients of the NNLL small-$x$ approximations 
(\ref{PgiNNL}) in $N$-space for the timelike gluon-gluon and gluon-quark 
splitting functions in QCD to the sixteenth order in $\as$.
\label{tab:pginnl}}
\end{center}
\end{table}
\begin{table}[p]
\vspace*{-0.1cm}
\small
\begin{center}
\begin{tabular}{|c|rrr|rrr|}\hline
 & & & & & & \\[-3mm]
 $n$ &
 $C^{\,(n)}_{\rm qg,1}$ & $C^{\,(n)}_{\rm qg,2}$ &
 $C^{\,(n)}_{\rm qg,3}$ & $C^{\,(n)}_{\rm qq,1}$ &
 $C^{\,(n)}_{\rm qq,2}$ & $C^{\,(n)}_{\rm qq,3}$ \\[2mm] \hline
 & & & & & & \\[-2mm]
$2$ &$-$7.0398681 & 3.1604938 & --      &$-$3.3757439 & --      & --   \\[1mm] 
$3$ & 0.2881972 & 12.609054 & 6.3209877 &$-$0.1122633 & 6.2624600 & -- \\[1mm]
$4$ & 3.8811194 & 19.180041 & 12.349337 & 1.4382426 & 9.0594422 & 2.2240512 \\[1mm]
$5$ & 6.2663008 & 23.451903 & 16.193141 & 2.4695470 & 10.673136 & 3.9805314 \\[1mm]
$6$ & 8.1028556 & 26.382647 & 18.550343 & 3.2699679 & 11.725549 & 5.2086911 \\[1mm]
$7$ & 9.6308924 & 28.524947 & 20.014717 & 3.9411132 & 12.479542 & 6.0662810 \\[1mm]
$8$ & 10.960497 & 30.185034 & 20.944287 & 4.5288220 & 13.063470 & 6.6786404 \\[1mm]
$9$ & 12.150652 & 31.537321 & 21.546983 & 5.0574954 & 13.545017 & 7.1280203 \\[1mm]
$10$& 13.236613 & 32.685054 & 21.945288 & 5.5417090 & 13.961930 & 7.4670190 \\[1mm]
$11$& 14.241194 & 33.691738 & 22.213216 & 5.9909234 & 14.336239 & 7.7296206 \\[1mm]
$12$& 15.180069 & 34.597838 & 22.396683 & 6.4116708 & 14.681305 & 7.9382263 \\[1mm]
$13$& 16.064511 & 35.429968 & 22.524875 & 6.8086793 & 15.005512 & 8.1079291 \\[1mm]
$14$& 16.902933 & 36.206143 & 22.616740 & 7.1854997 & 15.314277 & 8.2491157 \\[1mm]
$15$& 17.701801 & 36.938872 & 22.684797 & 7.5448791 & 15.611200 & 8.3690760 \\[1mm]
\hline
\end{tabular}
\vspace{1mm}
\caption{The corresponding N$^{\:\!3}$LL coefficients in (\ref{PqiNNL}) for the
timelike quark-gluon and quark-quark splitting functions in QCD to the 
sixteenth order in the strong coupling constant.
\label{tab:pqinnl}}
\end{center}
\vspace*{-4mm}
\end{table}



\newpage
\setcounter{equation}{0}
\section{Resummed coefficient functions for {\boldmath $F_T$}}
We now turn to the coefficient functions. For brevity, we will not discuss the
$\phi$-exchange case here (beyond the respective highest logarithms which are
directly related to those for $F_T$), as it will be of only theoretical 
interest in the near future. 
The corresponding results are included, however, in the {\sc Form} file of 
results distributed with the arXiv version of this article. 

The moments of the small-$x$ resummed terms of the transverse coefficient 
functions are
\beq
\label{CTiexp}
  c^{}_{T,\,\rm i}(N) \;=\; \sum_{n=1}^{\infty}\, \ar^{\,n}
  \left( \delta_{\rm \,ig\,}^{} c_{T\rm i,\,LL}^{\,(n)}(N)
   \:+\: c_{T\rm i,\,NLL}^{\,(n)}(N)
   \:+\: c_{T\rm i,\,NNL}^{\,(n)}(N)
   \:+\; \ldots \right) \; .
\eeq
The leading and next-to-leading logarithmic contributions for
$c_{T,\,\rm g}^{}$ can be written as
\beq
\label{cTgLL}
   c_{T\rm g,\,LL}^{\,(n)}(N) \;=\;
   \frac{(-4)^n\,\cf\, C_A^{\,n-1}}{(N-1)^{2n}}\: A^{(n)}_{T,\,\rm g}
\eeq

\vspace*{-6mm}
\noindent
and
\bea
\label{cTgNL}
   c_{T\rm g,\,NL}^{\,(n)}(N) \;=\;
   \frac{(-4)^n\,\cf\, C_A^{\,n-3}}{9(N-1)^{2n-1}}\:
   \Big[ - \cas\,  B^{\,(n)}_{T\rm g,1}
   \,+\, \ca \nf\, B^{\,(n)}_{T\rm g,2} 
   \,+\, 8\,\cf \nf\, B^{\,(n)}_{T\rm g,3} \Big] \; .
\eea
The coefficients in Eqs.~(\ref{PggLL}) and (\ref{PgqNL}) are given in Table
\ref{tab:cTgnl} analytically to the twelfth order in $\as$ (see the {\sc Form} 
file for the remaining four orders) and numerically for $n=13,\,\ldots\,16$.
In this case the general form and the generating function is obvious only for
the leading-logarithmic coefficients in Eq.~(\ref{cTgLL}) with \cite{series3}
\beq
  A^{\,(n)}_{T,\,\rm g} \;=\; \frac{2^{\,n}}{n!}\;
  \prod_{k=0^{\phantom{1}\!\!\!}}^{n-1} \; ( 4\:\!k + 1 )
\eeq
and
\beq
\label{cTg-cl}
  c_{T,\,\rm g}^{\;\rm LL}(N) \;=\; 
   \frac{C_F}{C_A}\: \left( c_{\phi,\rm g}^{\,T,\,\rm LL}(N) - 1 \right)
   \;=\; \frac{C_F}{C_A}\: \Bigg\{ \!\!
  \left( 1 + \frac{32\,\ca\, \ar}{(N-1)^2} \right)^{-1/4}\! - 1 \Bigg\} \; .
\eeq
%
%
Eq.~(\ref{cTg-cl}) agrees with the corresponding result of 
Ref.~\cite{ABKK11} for $c_{T,\,\rm g}^{}$ up to a factor of two arising from the
different normalization of this coefficient function already mentioned below 
Eq.~(\ref{cLg12}).

The relations for the quark coefficient functions corresponding to 
Eqs.~(\ref{cTgLL}) and (\ref{cTgNL}) can be cast in
the form
\beq
\label{cTqLL}
   c_{T\rm q,\,NLL}^{\,(n)}(N) \;=\;
   \frac{C_F}{C_A}\: c_{\phi\, q,\,\rm NLL}^{\,T\,(n)}(N) \;=\;
   \frac{(-4)^n\,\cf\nf\, C_A^{\,n-2}}{3(N-1)^{2n-1}}\: 4\,A^{(n)}_{T,\,\rm q}
\eeq
and
\beq
\label{cTqNL}
   c_{T\rm q,\,NNL}^{\,(n)}(N) \;=\;
   \frac{(-4)^n\,\cf\nf C_A^{\,n-4}}{3(N-1)^{2n-2}}\:
   \Big[ - \cas\,  B^{\,(n)}_{T\rm q,1}
   \,+\, \fr{8}{3}\: \ca \nf\, B^{\,(n)}_{T\rm q,2}
   \,+\, \fr{8}{3}\: \cf \nf\, B^{\,(n)}_{T\rm q,3} \Big] 
\eeq
with $n \geq 2$. The first sixteen coefficients in Eqs.~(\ref{cTqLL}) and 
(\ref{cTqNL}) can be found in Table \ref{tab:cTqnl}. Note the the faster
growth of these coefficients with $n$, as compared to the corresponding 
splitting function results in Tables \ref{tab:pginl} and \ref{tab:pqinl}, 
is largely (but only only) due to the different normalization in 
Eqs.~(\ref{cTgLL}) and Eqs.~(\ref{cTqLL}), which was employed to have mainly 
integer coefficient in Table \ref{tab:pginl}.

\begin{table}[thb]
\small
\vspace*{1mm}
\begin{center}
\begin{tabular}{|c|rrrc|}\hline
 & & & & \\[-3mm]
 $n$ &
 $A^{\,(n)}_{T,\,\rm g}$  & $B^{\,(n)}_{T\rm g,1}$ &
 $B^{\,(n)}_{T\rm g,2}$ & $B^{\,(n)}_{T\rm g,3}$ \\[2mm] \hline
 & & & & \\[-3mm]
 1  & 2          & $-$9         & --          & --                       \\[1mm]
 2  & 10         & 87/2         & --          & --                       \\[1mm]
 3  & 60         & 779          & 2           & 8                        \\[1mm]
 4  & 390        & 8620         & 67          & 115                      \\[1.5mm]
 5  & 2652       & 84224        & 1100        & $\FR{6193}{5}$           \\[3.5mm]
 6  & 18564      & 778449       & 14028       & $\FR{59811}{5}$          \\[3.5mm]
 7  & 132600     & 6974466      & 157500      & $\FR{765402}{7}$         \\[3.5mm]
 8  & 961350 	  & 61261449     & 1639437     & $\FR{13548231}{14}$     \\[3.5mm]
 9  & 7049900    & 530773430    & 16238552    & $\FR{176155814}{21}$     \\[3.5mm]
 10 & 52169260   & 4552643821   & 155338216   & $\FR{1505191630}{21}$    \\[3.5mm]
 11 & 388898120  & 38750254946  & 1448362604  & $\FR{140014436692}{231}$ \\[3.5mm]
 12 & 2916735900 & 327823740972 & 13242606390 & $\FR{391609950056}{77}$  \\[3.5mm]
 13 & {2.1987701\,10$^{10}\!$} & {2.7596825\,10$^{12}\!$} 
    & {1.1922955\,10$^{11}\!$} & {4.2417307\,10$^{10}\!$} \\[2mm]
 14 & {1.6647831\,10$^{11}\!$} & {2.3136533\,10$^{13}\!$} 
    & {1.0602610\,10$^{12}\!$} & {3.5208088\,10$^{11}\!$} \\[2mm]
 15 & {1.2652351\,10$^{12}\!$} & {1.9330232\,10$^{14}\!$} 
    & {9.3330885\,10$^{12}\!$} & {2.9110969\,10$^{12}\!$} \\[2mm]
 16 & {9.6474181\,10$^{12}\!$} & {1.6102477\,10$^{15}\!$} 
    & {8.1461913\,10$^{13}\!$} & {2.3992885\,10$^{13}\!$} \\[1.5mm]
\hline
\end{tabular}
\vspace{1.5mm}
\caption{The first sixteen $N$-space coefficients in Eqs.~(\ref{cTgLL}) -- 
(\ref{cTgNL}) for the LL and NLL small-$x$ approximations to the gluon 
coefficient function for the transverse fragmentation function.
\label{tab:cTgnl}}
\end{center}
\vspace*{-7mm}
\end{table}

\begin{table}[thb]
\footnotesize
\vspace*{1mm}
\begin{center}
\begin{tabular}{|c|cccc|}\hline
 & & & & \\[-3mm]
 $n$ &
 $A^{\,(n)}_{T,\,\rm q}$  & $B^{\,(n)}_{T\rm q,1}$ &
 $B^{\,(n)}_{T\rm q,2}$ & $B^{\,(n)}_{T\rm q,3}$ \\[2mm] \hline
 & & & & \\[-2mm]
 2  & 1                        & $-$1                          & --                           & --  \\[2mm]
 3  & $\FR{23}{3}$             & $\FR{340}{9}$                 & 1                            & -- \\[3.5mm]
 4  & $\FR{329}{6}$            & $\FR{4790}{9}$                & $\FR{133}{12}$               & $\FR{26}{3}$ \\[3.5mm]
 5  & $\FR{5884}{15}$          & $\FR{50917}{9}$               & $\FR{1604}{15}$              & $\FR{4211}{30}$ \\[3.5mm]
 6  & $\FR{14166}{5}$          & $\FR{2454268}{45}$            & $\FR{29869}{30}$             & $\FR{16313}{10}$ \\[3.5mm]
 7  & $\FR{144694}{7}$         & $\FR{157792304}{315}$         & $\FR{319122}{35}$            & $\FR{1743793}{105}$ \\[3.5mm]
 8  & $\FR{2130333}{14}$       & $\FR{156504164}{35}$          & $\FR{34659563}{420}$         & $\FR{13254173}{84}$ \\[3.5mm]
 9  & $\FR{71114144}{63}$      & $\FR{7404527591}{189}$        & $\FR{93174769}{126}$         & $\FR{7191782}{5}$ \\[3.5mm]
 10 & $\FR{530983954}{63}$     & $\FR{45735067426}{135}$       & $\FR{10344355237}{1575}$     & $\FR{20099736449}{1575}$ \\[3.5mm]
 11 & $\FR{43854388318}{693}$  & $\FR{861656350072}{297}$      & $\FR{111399799846}{1925}$    & $\FR{1925137106758}{17325}$ \\[3.5mm]
 12 & $\FR{110281846025}{231}$ & $\FR{769021780130564}{31185}$ & $\FR{10525291437281}{20790}$ & $\FR{3306988478369}{3465}$ \\[3.5mm]
 13 & {\small 3.6165904\,10$^{9}\!$} & {\small 2.0836819\,10$^{11}\!$}
    & {\small 4.4007722\,10$^{9}\!$} & {\small 8.1132902\,10$^{9}\!$} \\[2mm]
 14 & {\small 2.7496227\,10$^{10}\!$} & {\small 1.7521629\,10$^{12}\!$}
    & {\small 3.8036176\,10$^{10}\!$} & {\small 6.8422246\,10$^{10}\!$} \\[2mm]
 15 & {\small 2.0971243\,10$^{11}\!$} & {\small 1.4675034\,10$^{13}\!$}
    & {\small 3.2707410\,10$^{11}\!$} & {\small 5.7338592\,10$^{11}\!$} \\[2mm]
 16 & {\small 1.6039639\,10$^{12}\!$} & {\small 1.2249474\,10$^{14}\!$}
    & {\small 2.7996798\,10$^{12}\!$} & {\small 4.7804829\,10$^{12}\!$} \\[1.5mm]
\hline
\end{tabular}
\vspace{1.5mm}
\caption{ As Table \ref{tab:cTgnl}, but for the NLL and NNLL quark coefficient 
function in Eqs.~(\ref{cTqLL}) -- (\ref{cTqNL}).
\label{tab:cTqnl}}
\end{center}
\vspace*{-6mm}
\end{table}

As for the splitting functions, the next contributions to both transverse
coefficient functions are considerably more complex. Since the third-order
SIA coefficients functions have not been published so far, we give the 
third- and fourth-order quantities analytically. The higher orders are
presented numerically for $\ca = 3$ and $\cf = 4/3$ below. 
The third-order results are given by
\bea
c_{T,\,\rm g}^{\,(3)}(N) \!&\!=\!\!&
   -\; {64 \over (N-1)^6}\; 60\,\cas\cf
   \:+\: {64 \over (N-1)^5} \: \Big\{
      \fr{779}{9}\: \cas\cf - \fr{2}{9}\: \ca\cf\nf - \fr{64}{9}\: 
      \cf\nfs \Big\}
\\ & & \mbox{} \!\!
   \:-\: {64 \over (N-1)^4} \: \Big\{ \!\!
     \left( \fr{395}{12} - \fr{68}{3}\,\z2 \right)\, \cas\cf
     - \left( \fr{5}{3} + \fr{4}{3}\:\z2 \right)\, \ca\cfs 
     + \fr{67}{27}\: \ca\cf\nf - \fr{47}{9}\: \cf\nfs \Big\} 
     \;+\:\: \ldots  \!
\nn
\\[1mm]
c_{T,\rm q}^{\,(3)}(N) \!&\!=\!&
   -\; {64 \over (N-1)^5}\; \fr{92}{9}\:\ca\cf\nf
   \:+\: {64 \over (N-1)^4} \: \Big\{
      \fr{340}{27}\: \ca\cf\nf - \fr{8}{9}\:\cf\nfs \Big\}
\\ & & \mbox{} \!\!
   \:+\: {64 \over (N-1)^3} \: \Big\{
      \left( \fr{169}{324} - \fr{2}{9}\:\z2 \right)\, \ca\cf\nf
      - \left( \fr{1}{3} - \fr{8}{3}\:\z2 \right)\, \cfs\nf
      - \fr{2}{9}\: \cf\nfs \Big\} \;+\:\: \ldots \;\; . 
\nn
\eea

\pagebreak
\noindent
The expansions of the fourth-order transverse coefficient functions about
$N=1$ read
\bea
 c_{T,\,\rm g}^{\,(4)}(N) \!&\!=\!\!&
   {256 \over (N-1)^8}\; 390\,\cath\cf
   \:-\: {256 \over (N-1)^7} \: \Big\{
      \fr{8620}{9}\: \cath\cf - \fr{67}{9}\: \cas\cf\nf 
      - \fr{920}{9}\: \ca\cf\nfs \Big\}
\nn \\ & & \mbox{\hspn}
   \:+\: {256 \over (N-1)^6} \: \Big\{ \!\!
     \left( \fr{219007}{216} - 244\,\z2 \right)\, \cath\cf
     - \left( \fr{15}{4} + 18\,\z2 \right)\, \cas\cfs
     + \fr{451}{36}\:\cas\cf\nf 
\nn \\[-1mm] & & \mbox{\hspace*{2cm}}
     - \fr{5308}{27}\:\ca\cfs\nf + \fr{31}{27}\:\ca\cf\nfs 
     + \fr{44}{9}\: \cfs\nfs \Big\} \;+\:\: \ldots 
\eea
 and
\bea
 c_{T,\rm q}^{\,(4)}(N) \!&\!=\!\!&
   {256 \over (N-1)^7}\; \fr{658}{9}\,\ca\cf\nf
   \:-\: {256 \over (N-1)^6} \: \Big\{
      \fr{4790}{27}\: \cas\cf\nf - \fr{266}{27}\: \ca\cf\nfs
      - \fr{208}{27}\: \cfs\nfs \Big\}
\nn \\ & & \mbox{\hspn}
   \:+\: {256 \over (N-1)^5} \: \Big\{ \!\!
     \left( \fr{32423}{216} - \fr{80}{9}\:\z2 \right)\, \cas\cf\nf
     + \left( \fr{71}{9} - \fr{262}{9}\:\z2 \right)\, \ca\cfs\nf
     - \fr{958}{36}\:\cas\cf\nf
\nn \\[-1mm] & & \mbox{\hspace*{2cm}}
     - \fr{838}{81}\: \cfs\nfs + \fr{8}{9}\: \cf\nft
     \Big\} \;+\:\: \ldots \;\; .
\eea
 
\begin{table}[p]
\small
\begin{center}
\begin{tabular}{|c|rrr|rrr|}\hline
 & & & & & & \\[-3mm]
 $n$ &
 $C^{\,(n)}_{T\rm g,0}$ & $C^{\,(n)}_{T\rm g,1}$ &
 $C^{\,(n)}_{T\rm g,2}$ & $C^{\,(n)}_{T\rm q,1}$ &
 $C^{\,(n)}_{T\rm q,2}$ & $C^{\,(n)}_{T\rm q,3}$ \\[2mm] \hline
 & & & & & & \\[-2mm]
 2  & $\!\!-$0.0488460 &  --   & --    & $\!\!-$0.0411523 &       --  &       --  \\[1mm]
 3  & $\!\!-$0.0052813 & $\!\!-$0.0044582 & -- & $\!\!-$0.0543741 & $\!\!-$0.0020576 & -- \\[1mm]
 4  &    0.0648580 &    0.2598844 & 0.3689986 & 0.4087098 & 0.0190107 & 0.0329218 \\[1mm]
 5  &    0.1603366 &    0.7502113 & 1.1226934 & 1.1073150 & 0.0595507 & 0.0993878 \\[1mm]
 6  &    0.2804175 &    1.4552223 & 2.2807323 & 2.0085092 & 0.1178470 & 0.1996540 \\[1mm]
 7  &    0.4247389 &    2.3722220 & 3.8617425 & 3.1012694 & 0.1933078 & 0.3345043 \\[1mm]
 8  &    0.5931376 &    3.5017009 & 5.8830239 & 4.3806502 & 0.2857698 & 0.5048923 \\[1mm]
 9  &    0.7855611 &    4.8454857 & 8.3606018 & 5.8442114 & 0.3952535 & 0.7118052 \\[1mm]
 10 &    1.0020223 &    6.4059897 & 11.309349 & 7.4907736 & 0.5218662 & 0.9562145 \\[1mm]
 11 &    1.2425739 &    8.1858750 & 14.743115 & 9.3198698 & 0.6657575 & 1.2390556 \\[1mm]
 12 &    1.5072942 &    10.187889 & 18.674852 & 11.331466 & 8.2709784 & 1.5612246 \\[1mm]
 13 &    1.7962784 &    12.414785 & 23.116717 & 13.525807 & 1.0060672 & 1.9235757 \\[1mm]
 14 &    2.1096323 &    14.869274 & 28.080167 & 15.903318 & 1.2028488 & 2.3269235 \\[1mm]
 15 &    2.4474687 &    17.554012 & 33.576041 & 18.464554 & 1.4176262 & 2.7720455 \\[1mm]
 16 &    2.8099054 &    20.471582 & 39.614620 & 21.210152 & 1.6505807 & 3.2596851 \\[1mm]
\hline
\end{tabular}
\vspace{1.0mm}
\caption{The numerical coefficients of the third small-$x$ contributions 
(\ref{cTgNNL}) and (\ref{cTqNNL}) to the $N$-space gluon and quark coefficient 
functions for the fragmentation function $F_T$ to order $\as^{\,16}$.
\label{tab:cTinnl}}
\end{center}
\vspace*{-6mm}
\end{table}

\noindent
For the coefficients of the third logarithms in Table \ref{tab:cTinnl} we use
the notation
\bea
\label{cTgNNL}
   c_{T\rm g,\,NNL}^{\,(n)}(N) &\!=\!&
   {(-1)^n \over (N-1)^{2n-2}}\: \left(
     96^{\,n}\, C^{\,(n)}_{T\rm g,0} \,-\,
     96^{\,n-1}\, C^{\,(n)}_{T\rm g,1}\: \nf \,+\,
     96^{\,n-2}\, C^{\,(n)}_{T\rm g,2}\: \nfs
   \right) \; ,
\\[2mm]
\label{cTqNNL}
   c_{T\rm q,\,N^3L}^{\,(n)}(N) &\!=\!&
   {(-1)^n \over (N-1)^{2n-3}}\: \left(
     96^{\,n-1}\, C^{\,(n)}_{T\rm q,1}\: \nf \,-\,
     96^{\,n-1}\, C^{\,(n)}_{T\rm q,2}\: \nfs \,+\,
     96^{\,n-2}\, C^{\,(n)}_{T\rm q,3}\: \nft
   \right) \; . \quad
\eea

\begin{figure}[p]
\vspace*{-1mm}
\centerline{\epsfig{file=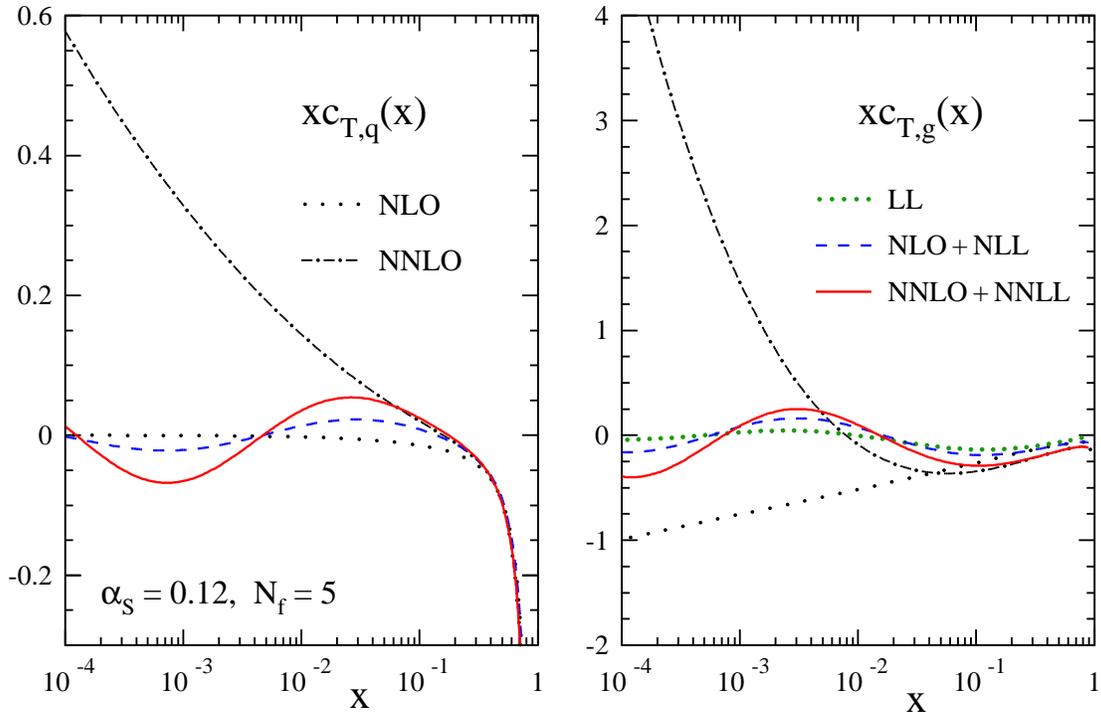,width=14.8cm,angle=0}\quad}
\vspace{-3mm}
\caption{ \label{fig:ctgexp2}
The quark and gluon coefficient functions for $F_T$ at a typical value of $\as$.
Shown are the NLO and NNLO fixed-order approximations, and the matched LL, NLL 
and NNLL
resummed results obtained (beyond LL) by adding the respective small-$x$ terms 
at all relevant higher orders.}
\vspace{-2mm}
\end{figure}

These results are illustrated in Fig.~\ref{fig:ctgexp2} for the same reference
point and $x$-range as in the previous section. The situation for 
$x\:\!c_{T,\,\rm g}^{}$ and $x\:\!c_{T,\,\rm q}^{}$ is largely analogous to that 
for the corresponding splitting functions $xP_{\rm gq}^{\,T}$ and 
$xP_{\rm qq}^{\,T}$ in Figs.~\ref{fig:Pgitexp2} and \ref{fig:Pqitexp2}.
The NLO and NNLO fixed-order approximations (the LO coefficient function  
$c_{T,\rm q}^{} = \delta(1-x)$ is obviously not visible in this figure) are
unreliable here from even larger $x$-values than above. The small-$x$ rise of 
the NNLO coefficient functions is removed by adding the NLL and NNLL 
resummations from order $\as^{\:\!3}$, leaving us with functions oscillating
about $xc_{T,k}^{} \approx 0$. The same behaviour, if with a considerably
smaller amplitude, can be established down to extremely small values of $x$
for the exactly known LL gluon coefficient function (\ref{cTg-cl}) already
determined in Ref.~\cite{ABKK11}. Also here it would be very interesting to
known one more order in $\as$ and the N$^3$LL resummation of 
$x\:\!c_{T,\,\rm g}^{}$. 
The latter, however, again requires (at least in the present framework) the 
calculation of the fourth-order contribution to the splitting function 
$P_{\rm gq}^{\,T}$.

It is instructive to briefly address the impact of the (scheme-independent) LL 
splitting functions (\ref{PLLsum}) and (scheme-dependent) LL coefficient 
functions, given in \MSb\ by Eq.~(\ref{cTg-cl}), on the scale dependence of the
fragmentation function $F_T$ and its `gluonic' counterpart $F_\phi$.
This is best done by considering the `timelike' physical evolution kernels 
$K_{ab}$ in Mellin space,
\beq
\label{KabPh}
  {d \over d \ln \Qs}
  \left( \begin{array}{c} \!\!F_T\!\! \\[0.5mm]
                          \!\!F_\phi\!\! \end{array} \right) 
  \;=\; 
  \left( \begin{array}{cc}
            \!  K_{\rm TT}     & \! K_{\rm T\phi} \!\! \\[1mm]
            \!  K_{\rm \phi T} & \! K_{\rm \phi\phi} \!\!
            \end{array} \right)
  \left( \begin{array}{c} \!\!F_T\!\! \\[0.5mm]
                          \!\!F_\phi\!\! \end{array} \right) \;\; ,
\eeq
which are given by the matrix elements of
\beq
\label{KmatPh}
  K \;\:=\;\: C\,P^{\,T} C^{\,-1} \:+\: 
  \beta(\ar)\: \frac{d\:\! C}{d \ar}\: C^{\,-1}
  \quad \mbox{ with } \quad
  C \;=\; \left( \begin{array}{cc}
            \!  c_{\,T,\rm q}^{}\!    & c_{\,T,\,\rm g}^{} \! \\[1mm]
            \!  c_{\,\phi,\rm q}^{}\! & c_{\,\phi,\rm g}^{} \!
           \end{array} \right)  \;\; ,
\eeq
and the splitting function matrix (\ref{PTmat}). In terms of powers of 
$(N-1)^{\,-1}$, the first term could be different from $P^{\,T}$ already at
leading logarithmic ($\,\as^{\,n} \,(N-1)^{\,-2n+2\,}$) accuracy.
However, the relations (\ref{PLLsum}) and Eq.~(\ref{cTg-cl}) imply 
\beq
\label{PCmatLL}
  P^{\,T}_{\,\rm LL} \;\:=\;
  \left( \begin{array}{cc}
            \!  0 & \fr{\cf}{\ca}\: P_{\!\rm gg,\,LL}^{\,T} \!\!\\[2mm]
            \!  0 & \;P_{\!\rm gg,\,LL}^{\,T} \!
           \end{array} \right)
  \quad \mbox{ and } \quad 
  C_{\rm LL}^{} \;=\; \left( \begin{array}{cc}
            \!  1 & \fr{\cf}{\ca}\: c_{\,\rm LL}^{} \!\!\\[2mm] 
            \!  0 & 1 + c_{\,\rm LL}^{} \!
           \end{array} \right) 
\eeq
with $c_{\,\rm LL}^{}$ given by the curly bracket in Eq.~(\ref{cTg-cl}).
Due to Eq.~(\ref{PCmatLL}) all such contributions to the matrix $K$ cancel, and
the factorization-scheme independent physical kernels are correctly given by
\beq
  K_{\rm TT,\,\rm LL}^{} \;=\; K_{\rm \phi T,\,\rm LL}^{} \;=\; 0
  \quad , \quad
  K_{\rm T\phi,\,\rm LL}^{} \;=\; P_{\!\rm gq,\,LL}^{\,T}
  \quad  , \quad  
  K_{\rm \phi\phi,\,\rm LL}^{} \;=\; P_{\!\rm gg,\,LL}^{\,T} \;\; .
\eeq
A study of the physical kernels (\ref{KabPh}) beyond the leading logarithmic
accuracy could be interesting, but is beyond the scope of the present article.

\setcounter{equation}{0}
\section{Resummed coefficient functions for {\boldmath $F_L$}}
Finally we briefly present the resummed results for the longitudinal 
fragmentation function $F_L$. Since the NNLO (third-order) coefficient 
functions for this observable are not yet known, only the respective two 
highest logarithms can be resummed for both the gluon and quark coefficient 
functions. 
The corresponding $N$-space expressions can be written as
\beq
\label{CLiexp}
  c^{}_{L,\,\rm i}(N) \;=\; \sum_{n=1}\, \ar^{\,n}
  \left( \delta_{\rm \,ig\,}^{} c_{L\rm i,\,LL}^{\,(n)}(N)
   \:+\: c_{L\rm i,\,NLL}^{\,(n)}(N)
   \:+\: c_{L\rm i,\,NNL}^{\,(n)}(N)
   \:+\; \ldots \right) \:\: ,
\eeq
with the gluon case given by
\beq
\label{cLgLL}
   c_{L\rm g,\,LL}^{\,(n)}(N) \;=\; -\:
   \frac{(-4)^n\,\cf C_A^{\,n-1}}{(N-1)^{2n-1}}\: A^{(n)}_{L,\,\rm g}
\eeq
and
\bea
\label{cLgNL}
   c_{L\rm g,\,NLL}^{\,(n)}(N) \;=\;
   \frac{(-4)^n\,\cf C_A^{\,n-3}}{9 (N-1)^{2n-2}}\:
   \Big[ \, \cas\, B^{\,(n)}_{L\,\rm g,1}
   \,-\, 9\,\ca \cf\, B^{\,(n)}_{L\,\rm g,2}
   \,-\, \ca \nf\, B^{\,(n)}_{L\,\rm g,3}
   \,-\, \cf \nf\, B^{\,(n)}_{L\,\rm g,4} \Big] \; .
\eea
As in the transverse case, the quark coefficient functions for $F_L$ are 
suppressed by one power of $\,\ln x\,$ or $\,(N-1)^{-1}$, but for $\,n>1$ take 
the otherwise analogous forms
\beq
\label{cLqLL}
   c_{L\,\rm q,\,NLL}^{\,(n)}(N) \;=\; -\:
   \frac{(-4)^n\,\cf\nf\, C_A^{\,n-2}}{3\:\!(N-1)^{2n-2}}\: A^{(n)}_{L,\,\rm q}
\eeq
and
\bea
\label{cLqNL}
   c_{L\,\rm q,\,NNL}^{\,(n)}(N) \;=\;
   \frac{(-4)^n\,\cf\nf\, C_A^{\,n-4}}{9 (N-1)^{2n-2}}\:
   \Big[ \, \cas\, B^{\,(n)}_{L\,\rm q,1}
   \,-\, \ca \cf\, B^{\,(n)}_{L\,\rm q,2}
   \,-\, \ca \nf\, B^{\,(n)}_{L\,\rm q,3}
   \,-\, \cf \nf\, B^{\,(n)}_{L\,\rm q,4} \Big] \; .
\eea
The coefficients in Eqs.~(\ref{cLgLL}) -- (\ref{cLqNL}) are given in 
Tables~\ref{tab:cLgnl} and \ref{tab:cLqnl}, as before giving the thirteenth to 
sixteenth order in a numerical form for brevity (the exact expressions can be
found in the {\sc Form} file distributed with this article). 
In this case the general formula is not even known for the LL coefficients 
which, like all other `unsolved' series above, involve unpleasantly large prime
numbers early in the expansion. For instance, the prime-factor decomposition of 
$A^{(7)}_{L,\,\rm g\,}$ reads $\,4\,\cdot\! 10691$.

\begin{figure}[p]
\vspace*{-5mm}
\centerline{\epsfig{file=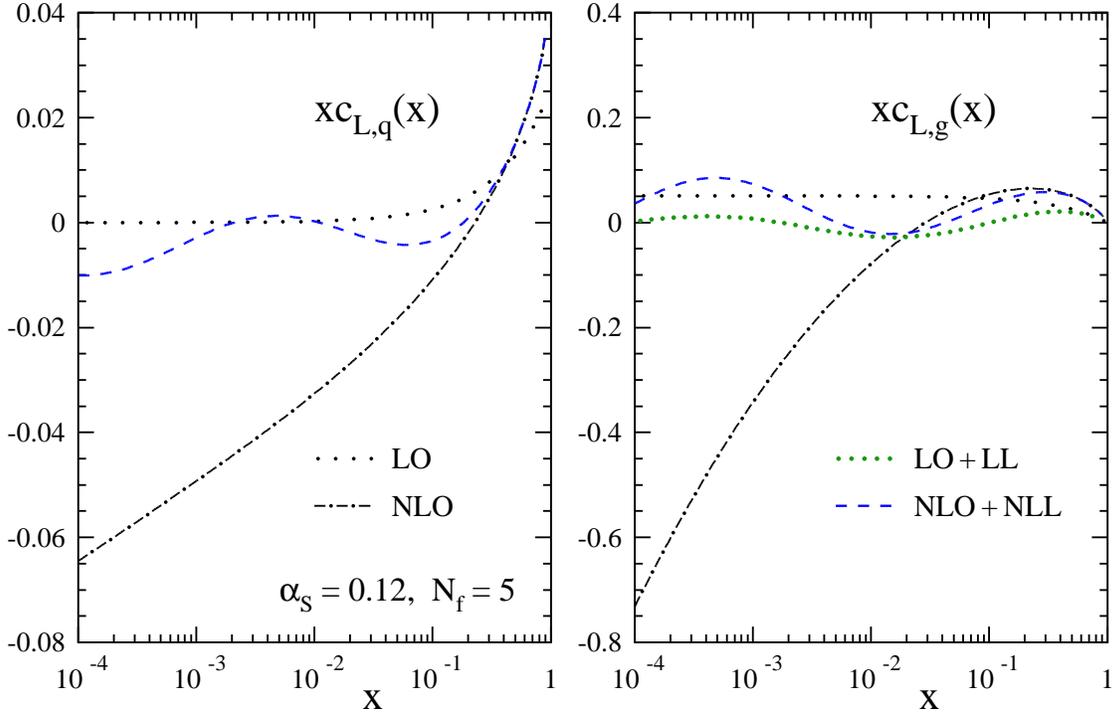,width=14.8cm,angle=0}\quad}
\vspace{-3mm}
\caption{ \label{fig:clgexp2}
The quark and gluon coefficient functions for $F_L$ at a typical value of $\as$.
Shown are the LO and NLO fixed-order approximations, and the matched LL (for 
$c_{L,\rm g\,}^{}$) and NLL resummed results obtained by adding the respective 
small-$x$ terms at all relevant higher orders.}
\vspace{-2mm}
\end{figure}

These results are illustrated in Fig.~\ref{fig:clgexp2} in the same manner as
those for $F_T$ in Fig.~\ref{fig:ctgexp2} above. While neither of the 
first-order (LO) coefficient functions includes any $\,x^{\,-1}\ln x\,$ terms
in the present case, also here the (now negative) small-$x$ spike of both 
second-order (NLO) coefficient functions is completely removed by adding the
corresponding all-order resummations of the small-$x$ logarithms, leaving
small oscillating functions with $xc_{L,p}^{}\approx 0$ at $x \lsim 10^{\,-2}$.

One may expect that the small-$x$ resummation of the longitudinal fragmentation
function will be the first to be extended to a higher accuracy as, in contrast
to the timelike splitting functions and the transverse fragmentation function
in the previous sections, `only' a third-order calculation is required for
deriving the NNLO$\,+\,$NNLL resummation. Note, however, that already the 
present results are sufficient for the corresponding resummation of the total
fragmentation function, obtained by integrating Eq.~(\ref{FrgFct}) over 
$\theta$, as the coefficient functions $c_{L,p}^{\,(n)}$ are suppressed by one 
power of $\,\ln x\,$ or $\,(N-1)^{-1}$ with respect to their transverse 
counterparts.

\begin{table}[thbp]
\footnotesize
\vspace*{1mm}
\begin{center}
\begin{tabular}{|c|rrrrc|}\hline
 & & & & & \\[-3mm]
 $n$ &
 $A^{\,(n)}_{L,\,\rm g}$  & $B^{\,(n)}_{L\,\rm g,1}$ &
 $B^{\,(n)}_{L\rm g,2}$   & $B^{\,(n)}_{L\rm g,3}$   & 
 $B^{\,(n)}_{L\rm g,4}$ \\[2mm] \hline
 & & & & & \\[-2mm]
 1  & 1         & 9            & --        & --         & --        \\[1.5mm]
 2  & 4         & 33           & 1         & --         & --        \\[1.5mm]
 3  & 22        & 723/2        & 5         & 3          & 30        \\[1.5mm]
 4  & 136       & 3530         & 30        & 56         & 376  \\[1.5mm]
 5  & 894       & 32447        & 195       & 722        & 3754 \\[2mm]
 6  & 6104      & 288590       & 1326      & 8000       & $\FR{172544}{5}$ \\[3.5mm]
 7  & 42764     & 2515565      & 9282      & 81722      & $\FR{1522436}{5}$ \\[3.5mm]
 8  & 305232    & 21633684     & 66300     & 793968     & $\FR{91820496}{35}$ \\[3.5mm]
 9  & 2209526   & 184263400    & 480675    & 7457476    & $\FR{779145058}{35}$ \\[3.5mm]
 10 & 16171672  & 1558144566   & 3524950   & 68371776   & $\FR{19627939136}{105}$ \\[3.5mm]
 11 & 119414516 & 13101831041  & 26084630  & 615603170  & $\FR{163580958068}{105}$ \\[3.5mm]
 12 & 888212208 & 109672261452 & 194449060 & 5465590416 & $\FR{14911681259824}{1155}$ \\[3.5mm]
 13 & {6.6468218\,10$^{9}\;$}  & {9.1464728\,10$^{11}\!$} 
    & {1.4583680\,10$^{9}\;$}  & {4.7987650\,10$^{10}\!$} 
    & {1.0652372\,10$^{11}\!$} \\[2mm]
 14 & {4.9997395\,10$^{11}\!$} & {7.6044089\,10$^{12}\!$} 
    & {1.0993851\,10$^{10}\!$} & {4.1752477\,10$^{11}\!$} 
    & {8.7587508\,10$^{11}\!$} \\[2mm]
 15 & {3.7774611\,10$^{11}\!$} & {6.3057119\,10$^{13}\!$} 
    & {8.3239155\,10$^{10}\!$} & {3.6055542\,10$^{12}\!$} 
    & {7.1815660\,10$^{12}\!$} \\[2mm]
 16 & {2.8649548\,10$^{12}\!$} & {5.2169677\,10$^{14}\!$} 
    & {6.3261758\,10$^{11}\!$} & {3.0939814\,10$^{13}\!$} 
    & {5.8747864\,10$^{13}\!$} \\[1.5mm]
\hline
\end{tabular}

\vspace{8mm}
\begin{tabular}{|c|ccccc|}\hline
 \phantom{16} & 
 \phantom{ 2.8649548\,10$^{12}\!$} & 
 \phantom{ 2.8649548\,10$^{12}\!$} & 
 \phantom{ 2.8649548\,10$^{12}\!$} & 
 \phantom{ 2.8649548\,10$^{12}\!$} & 
 \phantom{ 2.8649548\,10$^{12}\!$} \\[-2mm] 
 $n$ &
 $A^{\,(n)}_{L,\,\rm q}$   & $B^{\,(n)}_{L\,\rm q,1}$ &
 $B^{\,(n)}_{L\,\rm q,2}$  & $B^{\,(n)}_{L\,\rm q,3}$   &
 $B^{\,(n)}_{L\,\rm q,4}$ \\[2mm] \hline
 & & & & & \\[-2mm]
 2  & 2                          & $-9/2$                      &
      --                         & --                          &
      --        \\[2mm]
 3  & 12                         & 51                          &
      6                          & 4                           &
      --        \\[1.5mm]
 4  & $\FR{236}{3}$              & $\FR{1976}{3}$              &
      46                         & $\FR{122}{3}$               &
      $\FR{92}{3}$  \\[3.5mm]
 5  & $\FR{1610}{3}$             & $\FR{19777}{3}$             &
      329                        & 379                         &
      446 \\[3.5mm]
 6  & $\FR{56356}{15}$           & $\FR{915601}{15}$           &
      $\FR{11768}{5}$            & $\FR{17228}{5}$             &
      $\FR{14560}{3}$ \\[3.5mm]
\hline
\end{tabular}
\vspace{3mm}
\caption{Upper part: the first sixteen $N$-space coefficients in 
Eqs.~(\ref{cLgLL}) and (\ref{cLgNL}) for the LL and NLL small-$x$ 
resummed the gluon coefficient function for the longitudinal fragmentation 
function. \protect\linebreak
Lower part: the first five (NLL and NNLL) coefficients for the corresponding
quark coefficient function defined in Eqs.~(\ref{cLqLL}) and (\ref{cLqNL}).
\label{tab:cLgnl}}
\end{center}
\vspace*{-5mm}
\end{table}

\begin{table}[p]
\footnotesize
\vspace*{1mm}
\begin{center}
\begin{tabular}{|c|ccccc|}\hline
 & & & & & \\[-3mm]
 $n$ &
 $A^{\,(n)}_{L,\,\rm q}$   & $B^{\,(n)}_{L\,\rm q,1}$ &
 $B^{\,(n)}_{L\,\rm q,2}$  & $B^{\,(n)}_{L\,\rm q,3}$   & 
 $B^{\,(n)}_{L\,\rm q,4}$ \\[2mm] \hline
 & & & & & \\[-1.5mm]
 7  & $\FR{401944}{15}$          & $\FR{8167748}{15}$          &  
      $\FR{84996}{5}$            & $\FR{154428}{5}$            & 
      $\FR{236236}{5}$ \\[3.5mm]
 8  & $\FR{6784088}{35}$         & $\FR{499053868}{105}$       &  
      $\FR{868164}{7}$           & $\FR{28758068}{105}$        & 
      $\FR{45616904}{105}$ \\[3.5mm]
 9  & $\FR{148855862}{105}$      & $\FR{4294474801}{105}$      & 
      $\FR{6390999}{7}$          & $\FR{84243073}{35}$         & 
      $\FR{135167864}{35}$ \\[3.5mm]
 10 & $\FR{3295405924}{315}$     & $\FR{36585726017}{105}$     & 
      $\FR{142228288}{21}$       & $\FR{734599784}{35}$        & 
      $\FR{2115778496}{63}$ \\[3.5mm]
 11 & $\FR{24496904632}{315}$    & $\FR{103121715842}{35}$     &  
      $\FR{1061967908}{21}$      & $\FR{286309749296}{1575}$   & 
      $\FR{151010702344}{525}$ \\[3.5mm]
 12 & $\FR{2015422894136}{3465}$ & $\FR{85829821660568}{3465}$ & 
      $\FR{87708776636}{231}$    & $\FR{9038620655308}{5775}$  & 
      $\FR{6030487800584}{2475}$ \\[3.5mm]
 13 & {4.3730248\,10$^{9}\;$}  & {2.0729526\,10$^{11}\!$} 
    & {2.8644635\,10$^{9}\;$}  & {1.3405311\,10$^{10}\!$} 
    & {2.0469392\,10$^{10}\!$} \\[2mm]
 14 & {3.3024706\,10$^{10}\!$} & {1.7282691\,10$^{12}\!$} 
    & {2.1699543\,10$^{10}\!$} & {1.1428789\,10$^{11}\!$} 
    & {1.7086228\,10$^{11}\!$} \\[2mm]
 15 & {2.5036749\,10$^{11}\!$} & {1.4363931\,10$^{13}\!$} 
    & {1.6497736\,10$^{11}\!$} & {9.7038944\,10$^{11}\!$} 
    & {1.4189917\,10$^{12}\!$} \\[2mm]
 16 & {1.9045398\,10$^{12}\!$} & {1.1906480\,10$^{14}\!$} 
    & {1.2582746\,10$^{12}\!$} & {8.2093343\,10$^{12}\!$} 
    & {1.1736262\,10$^{13}\!$} \\[1.5mm]
\hline
\end{tabular}
\vspace{1.5mm}
\caption{Continuation of the part Table \ref{tab:cLgnl} for
the quark coefficient function for $F_L$ to order $\as^{\:\!16}$.
\label{tab:cLqnl}}
\end{center}
\vspace*{-6mm}
\end{table}
 
\setcounter{equation}{0}
\section{Summary and Outlook}
We have derived the all-order resummation of the highest three small-$x$ double 
logarithms, 
\beq
\label{sum-logs}
  \as^{\,n}\, x^{\,-1} \ln^{\,2n-\ell_0-\ell\!}x\, 
 \quad \mbox{ with } \quad \ell \;=\; 0,\:1,\:2 \; ,
\eeq 
for all four flavour-singlet timelike splitting functions -- with $\ell_0 = 2$ 
for $P_{\rm gq}^{\,T}$ and $P_{\rm gg}^{\,T}$ and $\ell_0 = 3$ for 
$P_{\rm qq}^{\,T}$ and $P_{\rm qg}^{\,T}$ -- 
and for both singlet coefficient functions for the transverse fragmentation 
function $F_T$ in semi-inclusive electron-positron annihilation (SIA) -- with 
$\ell_0 = 2$ for $c_{T,\,\rm q}^{}$ and $\ell_0 = 1$ for $c_{T,\,\rm g}^{}$~-- 
together with the corresponding results for SIA via an intermediate scalar
$\phi$ like the Higgs boson in the heavy top-quark limit. 
For the longitudinal fragmentation function $F_L$ present fixed-order results, 
which serve as input quantities for the resummation, allow only the 
determination of the highest two logarithms, i.e., $\ell = 0,\,1$ in 
Eq.~(\ref{sum-logs}) with $\ell_0 = 3$ for $c_{L,\,\rm q}^{}$ and $\ell_0 = 2$ 
for $c_{L,\,\rm g}^{}$. 
 
The coefficients of the above logarithms have been calculated explicitly to 
order $\as^{\,16}$ which is not the highest computationally feasible order, but 
sufficient for numerically accurate results down to $x = 10^{\,-4}$, a range
in $x$ that should be more than sufficient for all foreseeable analyses of data.
These calculations have been performed in Mellin-$N$ space, using the latest
versions of {\sc Form} and {\sc TForm} \cite{Form3,TForm} at all stages.
The results agree with the leading logarithmic (LL) result of 
Refs.~\cite{LLxto0} for the splitting functions $P_{\rm gq}^{\,T}$ and 
$P_{\rm gg}^{\,T}$, and with the only additional result so far derived in the 
\MSb\ scheme, the recent LL contributions to the coefficient function 
$c_{T,\,\rm g}^{}$ \cite{ABKK11}.

The resummation has been derived by decomposing the unfactorized 
partonic fragmentation functions $\widehat{F}_{a, p}(x,\as,\ep)$ in dimensional
regularization at any order $\as^{\,n}$ into $n$ (or $n-1$ in the quark cases) 
contributions of the form
\beq
\label{sum-Ddim}
  \ep^{\,-2\,n+n_0^{}}\, x^{\,-1-2\:\!k\,\ep}\, 
  ( A \,+\, B\,\ep \,+\, C\,\ep^2 \,+\: \ldots\, )
  \quad \mbox{ with } \quad k \;=\; 1,\:2,\: \ldots,\: n
\eeq
and $\,n_0^{} = 1\,$ for $\,a = T,\, \phi$ and $p = \rm g\,$, 
$\,n_0^{} = 2\,$ for $\,a = T,\, \phi$ and $p = \rm q\,$ and for 
$\,a,p = L,\rm g\,$, and $\,n_0^{} = 3\,$ for $\,a,p = L,\,\rm q\,$, with
the $k=1$ contributions missing in the quark cases. The KLN-related 
cancellations between the contributions in Eq.~(\ref{sum-Ddim}), together with
the powers of $\ep$ fixed by fixed-order calculations 
\cite{NLOcoeff1,NLOcoeff2,RvN2loop,MM06,MMV06,PT1loop,CFP80,KKST80,FKL81,%
MOKK01,MV2,AMV1}, lead to overconstrained systems of equations for the 
leading logarithmic, next-to-leading logarithmic (NLL) [and next-to-next-to-%
leading logarithmic (NNLL)] expansion parameters $A$, $B$ [and $C$] in the
decomposition (\ref{sum-Ddim}) which can be solved to (in principle) any 
order~$n$.
Given the large number of extra constraints and checks -- including the correct
predictions of the respective highest two small-$x$ logarithms in the 
third-order timelike splitting functions \mbox{\cite{MV2,AMV1}} and the 
non-trivial all-order agreement with the known LL results \cite{LLxto0,ABKK11} 
-- there is no need for an additional derivation of the decomposition 
(\ref{sum-Ddim}) from the structure of higher-order Feynman diagrams and 
phase-space integrations.

Whilst the setup of the resummation is elegant and simple, most of the new
results are not, as we have not succeeded to find the general expressions and
generating functions for the resulting series of coefficients, with the 
exception of the NLL corrections to the splitting functions $\,C_F^{\,-1}\, 
P_{\rm gq}^{\,T}$ and $P_{\rm gg}^{\,T}$ in the limit $\cf = 0\,$. 
The results have therefore been presented via detailed $N$-space tables which, 
hopefully, will be used for finding some of the now unknown general expressions.
The most interesting target in this respect are the non-integer coefficients in 
Table~\ref{tab:pginl}, as the solution of any one of these three series would
be sufficient to clarify the analytic structure of all NLL 
$(\as^{\,n}\, x^{\,-1} \ln^{\,2n-3\!}x\,)$ contribution to the matrix of the 
timelike splitting functions.

The small-$x$ resummation has a striking effect on the numerical behaviour of
the splitting functions and coefficient functions in the region 
$x \lsim 10^{\,-2}$. All fixed-order spikes for $x \ra 0$, which dwarf their
single-logarithmic counterparts in the spacelike splitting functions and 
deep-inelastic scattering (DIS) \cite{BFKL1,BFKL2,BFKL3,CH94,NL-BFKL}, are 
removed by forming the N$^n$LO$\,+\,$N$^n$LL combinations of fixed-order and
higher-order resummed results, mostly leaving small and apparently oscillating
functions. This behaviour is qualitatively similar to the LL results of 
Ref.~\cite{LLxto0,ABKK11} which are known in a closed form and thus can be 
evaluated down to extremely small values of $x$. While some theoretical 
questions remain that can only be clarified by future third- and fourth-order
calculations, the present resummation should prove sufficient for analyses of 
SIA data in the foreseeable future.

We have verified that the present approach can be extended to the 
non-$x^{\,-1}$ double logarithms in the (even-$N$ based) DIS structure 
functions $F_2$ and $F_L$ (recall that there are no `genuine' $x^{\,-1}$ double
logarithms in DIS; those encountered in the $\phi$-exchange coefficient 
functions in Refs.~\cite{SMVV1,DGGL} are artifacts of using the heavy-top 
approximation outside its domain of validity). 
These double-logarithmic terms form the leading small-$x$ contributions in the 
non-singlet cases, see Refs.~\cite{SmallxNS} for the LL resummation of the 
spacelike non-singlet splitting functions; they can be relevant at intermediate
values of $x$ also in flavour-singlet quantities, see Ref.~\cite{MVV6}. The 
corresponding NNLL resummations will be presented in a subsequent publication.

One may expect that, analogous to the large-$x$ cases in Refs.~\cite{SMVV1,MV5},
the resummation of the small-$x$ double logarithms can be extended to (all) 
higher powers of the prefactor $x$ in Eq.~(\ref{sum-logs}) for the quantities 
considered here (and their even-$N$ spacelike counterparts) 
-- but not for the asymmetric fragmentation function $F_A$ which is related to 
the odd-$N$ structure function $F_3$ known to receive additional contributions 
with $1/n_c$ and higher group factors \cite{SmallxNS,MVV10}.
We have explicitly checked the direct generalization of our approach to the LL 
and NLL $x^{\,a}$ contributions in singlet SIA for $a = 0,\,\ldots,\:\!6$. 
It works, but only for $a = 0$ and even values, and with the form 
(\ref{sum-Ddim}) 
replaced~by
\beq
\label{sum-Ddim2}
  \ep^{\,-2\,n+1}\, x^{\,a-k\,\ep}\,
  ( A \,+\, B\,\ep \,+\, C\,\ep^2 \,+\: \ldots\, )
  \quad \mbox{ with } \quad k \;=\; \:2,\: \ldots,\: n+1 
\eeq
which, in fact, is what one may have `naively' expected from 
Refs.~\cite{RvN2loop} also for the $x^{\,-1}$ terms. \linebreak
The predictions resulting from Eq.~(\ref{sum-Ddim2}) should be useful in the 
context of future third- and fourth-order calculations. 
Conceivably also all small-$x$ double logarithms in the timelike and spacelike 
higher-order singlet splitting functions (and the corresponding SIA and DIS 
coefficient functions) could turn out to be `inherited' from lower-order 
quantities.  This issue deserves further studies including the case of 
${\cal N}\!=\! 4$ Super Yang-Mills theory addressed, for example, in 
Ref.~\cite{BK06,DokM06}.

A {\sc Form} file of our results presented in Sections 3 -- 5 can be obtained 
by downloading the source of this article from the {\tt arXiv} servers or from 
the author upon request.

\vspace*{5mm}
\noindent{\bf Acknowledgments}
 
\noindent
I am grateful to Sven Moch for critically reading the manuscript of this 
article.  It is a pleasure to thank Jos Vermaseren for providing huge 
efficiency improvements for several {\sc Form} programs used to compute and 
mass-factorize the resummed expressions. 
Eq.~(\ref{Bgq2qg3}) was found after discussing Table \ref{tab:pginl} with my 
Liverpool colleague John Gracey, who spotted that the difference on the 
r.h.s.~leads to a simpler series.
This research has been supported by the UK Science \& Technology Facilities 
Council (STFC) under grant number ST/G00062X/1. The author is also a member
of the European-Union funded network {\it LHCPhenoNet} with contract number 
PITN-GA-2010-264564.

\end{document}